\documentclass[manuscript,nonacm]{acmart}
\renewcommand\footnotetextcopyrightpermission[1]{}

\usepackage{booktabs} 
\usepackage{array}
\usepackage{multirow}
\usepackage{enumitem}
\usepackage{soul}
\usepackage{colortbl}
\usepackage{wasysym}
\usepackage{subfig}
\usepackage{float}

\usepackage{multirow}
\usepackage{makecell}
\usepackage{longtable,booktabs}

\newcommand{\tool}{ParPal}

\begin{document}
\title{``Learning Together'': AI-Mediated Support for Parental Involvement in Everyday Learning }

\author{Yao Li}
\affiliation{%
  \institution{College of Information Sciences and Technology, The Pennsylvania State University}
  \city{State College}
  \state{Pennsylvania}
  \country{United States}
}

\author{Jingyi Xie}
\affiliation{%
  \institution{Department of Industrial and Systems Engineering, San José State University}
  \city{San José}
  \state{California}
  \country{United States}
}

\author{Ya-Fang Lin}
\affiliation{%
  \institution{College of Information Sciences and Technology, The Pennsylvania State University}
  \city{State College}
  \state{Pennsylvania}
  \country{United States}
}

\author{He Zhang}
\affiliation{%
  \institution{College of Information Sciences and Technology, The Pennsylvania State University}
  \city{State College}
  \state{Pennsylvania}
  \country{United States}
}

\author{Ge Wang}
\affiliation{%
  \institution{Siebel School of Computing and Data Science, University of Illinois at Urbana-Champaign}
  \city{Champaign}
  \state{Illinois}
  \country{United States}
}

\author{Gaojian Huang}
\affiliation{%
  \institution{Department of Industrial and Systems Engineering, San José State University}
  \city{San José}
  \state{California}
  \country{United States}
}

\author{Rui Yu}
\affiliation{%
  \institution{Department of Computer Science and Engineering, University of Louisville}
  \city{Louisville}
  \state{Kentucky}
  \country{United States}
}

\author{Si Chen}
\affiliation{%
  \institution{University of Notre Dame}
  \city{Notre Dame}
  \state{Indiana}
  \country{United States}
}

\begin{abstract}
Family learning takes place in everyday routines where children and caregivers read, practice, and develop new skills together. Although AI is increasingly present in learning environments, most systems remain child-centered and overlook the collaborative, distributed nature of family education. This paper investigates how AI can mediate family collaboration by addressing tensions of coordination, uneven workloads, and parental mediation. From a formative study with families using AI in daily learning, we identified challenges in responsibility sharing and recognition of contributions. Building on these insights, we designed FamLearn, an LLM-powered prototype that distributes tasks, visualizes contributions, and provides individualized support. A one-week field study with 11 families shows how this prototype can ease caregiving burdens, foster recognition, and enrich shared learning experiences. Our findings suggest that LLMs can move beyond the role of tutor to act as family mediators -- balancing responsibilities, scaffolding intergenerational participation, and strengthening the relational fabric of family learning.
\end{abstract}



\authorsaddresses{}
\maketitle
\section{Introduction}

Family learning encompasses the everyday ways children and caregivers engage in learning together -- from reading stories, practicing sports, and exploring cultural traditions to supporting homework. Unlike classroom learning, which follows a formal curriculum, family learning is embedded in daily routines and relationships. It involves not only the transfer of knowledge but also the coordination of responsibilities, negotiation of values, and recognition of each member’s contributions \cite{falk2000learning,rogoff2003cultural}. Educational research highlights the importance of parental involvement in children’s learning. For example, Epstein’s framework of six types of involvement identifies ``learning at home’’ as a crucial domain, where parents and caregivers actively support academic activities such as homework, reading, and skill-building \cite{epstein2018school}. This perspective emphasizes that parents’ roles extend beyond simply checking correctness: they structure learning processes, model persistence, and help children connect schoolwork with broader life experiences. Family learning, therefore, is not merely about content delivery, but also about relational, emotional, and organizational practices that shape children’s development \cite{moll1992funds,hochschild1983managedheart}.

HCI research has long studied family collaboration and coordination, showing how calendars, to-do lists, and communication platforms help households manage everyday life \cite{neustaedter2009calendar, davidoff2010routine}. Within learning contexts, prior work highlights both opportunities and challenges. On the positive side, collaboration among caregivers can enrich children’s experiences by bringing in diverse expertise and perspectives, sustaining motivation, and strengthening family bonds \cite{falk2000learning,kim2025bridging}. At the same time, challenges are well documented: divisions of labor are often uneven, coordination between multiple caregivers can be minimal, and invisible work -- such as planning or emotional scaffolding -- goes unrecognized \cite{reitberger2009tasks,jo2020parentingstress}. Research also points to tensions when parents’ educational values or teaching approaches diverge, creating conflicts or inconsistencies in support \cite{lin2024coparenting}.

Recently, large language models (LLMs) and AI tutors have been promoted as personalized educational assistants, promising instant answers, practice problems, and scaffolding strategies. Tools such as Khanmigo (Khan Academy) \cite{khanmigo2025} or Duolingo’s AI tutor \cite{duolingo2025} position LLMs as one-on-one companions that guide students step by step through assignments. However, most of these systems are designed for either the child alone or a teacher, overlooking the distributed, collaborative, and relational nature of family learning. By framing homework as a site of ``learning at home’’ and shared responsibility, we argue that AI systems must also consider how to support collaboration among multiple caregivers, balance workloads, and make hidden contributions more visible.

The dynamics of family learning are particularly salient in China, where homework plays an outsized role in children’s education. Driven by competitive entrance exams and a strong cultural emphasis on academic success \cite{gao2025homeworkwars}, families often invest significant time and energy in tutoring at home. Parents and grandparents frequently share responsibilities, dividing subjects based on expertise or availability. This intensive involvement creates opportunities for intergenerational learning but also amplifies challenges of uneven workload, limited coordination, and concerns over children’s dependence on external support. In this context, AI tutoring tools are rapidly becoming part of everyday routines. Popular platforms such as Zuoyebang and Xiaoyuan Souti automatically solve homework problems and provide practice questions \cite{zuoyebang2025}, while newer LLM-based apps like Quark Gaokao use exam data to recommend study pathways \cite{quarkgaokao2024}. These trends make China a critical site for examining both the potential and pitfalls of AI in family learning.

Through a formative study with families who incorporated AI into activities such as homework tutoring, cultural literacy, and skill-building, we identified challenges of limited coordination, imbalanced responsibilities, and parental gatekeeping. Building on these insights, we designed \textit{FamLearn}, an AI-augmented system that distributes learning tasks across caregivers, provides individualized support, and surfaces hidden contributions. A one-week field study with 11 families demonstrates how such a system can not only ease caregiving burdens but also foster recognition, shared accountability, and richer family learning experiences. Our work makes three main contributions. First, we present the design and one-week deployment of \textit{FamLearn}, a system that suggests and distributes at-home learning tasks across caregivers, provides individualized support, and makes hidden contributions visible. Second, we extend theories of family learning and parental involvement \cite{epstein2018school,moll1992funds} by reframing “learning at home’’ as a collaborative, intergenerational practice rather than a dyadic parent–child activity. Third, we demonstrate how AI systems can surface invisible caregiving work \cite{hochschild1983managedheart,reitberger2009tasks}, rebalance responsibilities, and calibrate trust between human and AI roles. Building on these contributions, we discuss how AI can help shift from child-centered or parent-centered approaches toward family- and education-centered collaboration.

\section{Related Works}

\subsection{Family Collaboration and Coordination}
Prior HCI research has extensively explored how families coordinate and collaborate for daily activities, tasks, and family events. These activities include asking for and offering assistance with errands \cite{sohn2012examination}, and managing essential household documentation \cite{sannon2020doing}. Particularly, some domestic artifacts have been designed to support this coordination among caregivers, ranging from calendars \cite{neustaedter2009calendar, neustaedter2006linc}, to-do lists \cite{reitberger2013invisible}, clocks \cite{brown2007locating}, and refrigerators \cite{taylor2006augmenting}, supporting scheduling, planning and creating mutual awareness for task coordination within family members. 
Beyond physical artifacts, mobile technologies have become integral to facilitating family communication and coordination \cite{neustaedter2007digital, davidoff2010routine, pan2015family, schaefbauer2015snack}. For instance, \citet{davidoff2010routine} highlight how routine structures, when supported by technologies like calendars and reminders, can better support intra-family collaboration and be more easily aware of changes in daily routines. Similarly, \citet{reitberger2013invisible} also emphasize the importance of making invisible labor visible and fostering appreciation to promote fairness in task allocation. Researchers have also examined how parents collaborate to complete a specific caregiving activity. Prior work has explored parental collaboration in domains such as managing children’s sleep routines \cite{shin2022more}, coordinating school drop-offs and pick-ups \cite{davidoff2010routine}, and reaching agreement on medical decisions \cite{kirchner2020just, song2018bebecode}. Additionally, some studies also reveal the diverse needs of families, such as cross-cultural parents \cite{yarosh2016best}, divorced parents \cite{odom2010designing, dworkin2016coparenting} or parents with kids having special needs \cite{grant2016intervention, nikkhah2022family}. For instance, \citet{grant2016intervention} explore how parents of children with cancer adapt communication and caregiving roles across phases of treatment. However, while these studies mainly focus on the coordination of household activities, fewer have addressed how such coordination methods extend to the daily family learning context, where tasks like reading, practicing skills, and completing homework also need coordination among caregivers.

\subsection{AI Supported Caregiver-Child Interaction} 
A growing body of HCI literature has investigated how AI tools can support learning activities between caregivers and children at home. Epstein’s Six Types of Involvement framework identifies six partnerships, including parenting, communication, volunteering, learning at home, decision-making and collaborating with community \cite{epstein2018school}. Particularly, learning at home provides information about how families help their child learn at home in homework and other interactive activities. Two main research directions have emerged, including using AI to enhance caregiver-child activities such as reading and storytelling, and exploring how AI can help caregivers share responsibilities, especially in collaborative or creative tasks.

Specifically, in interactive reading and storytelling, systems such as \textit{StoryBuddy} allow AI to take part in caregiver-child reading sessions, with features such as generating questions, as well as providing hints and discussion topics \cite{zhang2022storybuddy}. Caregivers may either join the activity or allow the AI to lead the activity when they are not available, and review their child’s reading and interaction afterwards, keeping them involved even when not present in real time. Additionally, AI systems can provide more personalized support based on a child’s interests and reading levels. Caregivers can then provide input such as deciding what to ask and how to respond to AI systems (e.g., \textit{StoryMate} \cite{chen2025characterizing}). In these human-AI systems, caregivers are still able to adjust or override what the AI produces, to minimize AI risks of producing inappropriate interactions. Involving caregivers in learning with AI systems can benefit both relationship building and practical time management among caregivers and children.

In more collaborative settings, studies have also explored how AI can help caregivers of different generations or roles contribute to children’s learning process. In Kim et al.’s study~\cite{kim2025bridging}, grandparents and children co-created stories using an AI-assisted tool that supported brainstorming and planning. In this process, AI divided the work, kept track of ideas, and helped participants understand each other’s input. These features may make each caregiver’s effort in child learning more visible and help rebalance roles of caregivers with different technical abilities or availability. Furthermore, with more capabilities, AI systems are expected to be involved in more than simply learning support, but also potentially in emotional and relational challenges during the child learning process. For example, a recent study of 78 families recorded around 475 hours of caregiver-child homework sessions and used AI to analyze emotions and conflict patterns \cite{gao2025homework}, and found that even when caregivers were trying to be supportive in children’s homework activity, caregivers might still have conflicts with children due to varied expectations, misunderstandings, or stress. These findings highlight that the design of AI systems that focus on family learning support may focus on both content delivery and the emotional dynamics of caregivers.

\subsection{AI for Family Learning}
Recent years have seen the growing integration of AI into home learning environments, with both commercial tools and research prototypes offering various levels of support for families. Commercial platforms such as Zuoyebang and Xiaoyuan Souti~\cite{zuoyebang2025,xiaoyuandouti2025} reduce parental workload by providing instant answers, photo-based search, and personalized practice. More advanced systems like Khanmigo~\cite{khanmigo2025} and ChatGPT's ``Study and Learn'' feature emphasize structured guidance and promote self-directed learning. These tools primarily target children as end-users, though recent findings show that parents are actively introducing LLMs into everyday learning interactions, such as in language learning~\cite{duolingo2025}, cultural exploration~\cite{quan2025parents}, and even mental health support~\cite{elwahsh2025linguistic}. In parallel, research efforts are exploring how families can learn about AI together, as in family-based AI literacy initiatives~\cite{long2022family,druga2022family,han2025empowering}, and how conversational agents or robots can scaffold family interactions around educational content~\cite{zhang2022storybuddy,ho2025empowering,liu2025bricksmart}.

In the research domain, an emerging body of work focuses on AI systems that explicitly support collaborative or parent-mediated learning. For instance, hybrid tutoring systems combine LLM-based generative feedback with curriculum-aligned scaffolding to assist caregivers in math support~\cite{venugopalan2025combining,nguyen2024designing}. Robot-mediated learning environments have also explored flexible parental involvement, empowering caregivers to review or control AI-generated content~\cite{ho2025set,sun2024exploring,rosman2025exploring}. These tools are particularly valuable in reducing anxiety or effort among caregivers who feel underprepared~\cite{shang2025learner,ho2025empowering}. Yet, despite their promise, challenges persist. Studies have documented tensions between safety and autonomy~\cite{kurian2025designing}, gaps in parental literacy or trust in AI~\cite{quan2024young,eira2025parents,perella2024parent}, and sociotechnical issues such as content appropriateness, personalization, and transparency~\cite{yang2025home,yu2025safeguarding,seo2025enhancing}. As families increasingly adopt AI tools into home routines~\cite{han2024teachers,gao2025homework}, the need to design for intergenerational participation, flexible moderation, and co-learning becomes more evident.

However, most existing AI systems still treat family learning as either a child-centered or parent-centered activity, rather than a collaborative, distributed, and relational process. Many tools overlook the invisible coordination work among caregivers or assume a singular caregiver-child dyad~\cite{zhang2022storybuddy,shang2025learner}. While some efforts consider parental scaffolding or emotional labor~\cite{ho2025set,quan2025parents}, few aim to systematically surface or rebalance contributions across family members. Our work addresses this gap by reframing family learning as a dynamic collaboration among multiple caregivers and exploring how AI systems can support this complexity. We build on prior literature to propose design mechanisms that support equitable task distribution, foster recognition, and calibrate trust in both human and AI roles. In doing so, we extend existing research from individualized tutoring and AI literacy to holistic, family-centered AI support.

\section{Formative Study}
We first conducted a formative study to understand (1) families' experiences with daily family learning, (2) their use cases of AI tools to support this process, and (3) their expectations of AI-empowered learning tools, through which to mitigate existing challenges and identify design opportunities. 

\subsection{Method} 
We recruited 12 adult caregivers from 7 families through our prior contact and snowball sampling. Eligibility criteria required that participants have a child enrolled in elementary or junior high school, experience with tutoring, and prior use of AI tools to facilitate family learning. 
If a participant shares tutoring responsibilities with other family members, we invited all available and willing caregivers from the same household to attend. 
Participation was voluntary, and no compensation was provided.

We conducted one-on-one semi-structured interviews with questions about 
(1) family background such as grade level of children;  
(2) family learning practices and collaboration between family members if more than one caregiver is involved; 
(3) experience with using AI tools in family learning, including use scenarios, benefits and challenges; and
(4) expectations of AI tools for family learning. 
Each interview lasted for 60 minutes and was audio-recorded after consent. 
Two authors conducted inductive thematic analysis~\cite{braun2006using}, iteratively open coding the transcribed interview data and grouping codes into themes. All authors reviewed and finalized these themes during weekly meetings. 
We identified four Key Insights (KI) regarding current family learning practices, as well as caregivers' expectations that are insufficiently supported by existing AI-powered tutoring tools.

\subsection{Key Insights}
\label{formative_insights}

\paragraph{\textbf{KI1: Subject-based or availability-based tutoring involving multiple caregivers.}}
Five out of seven families indicated that more than one caregiver in their household is involved in everyday family learning, most commonly parents and grandmothers. 
In some households, the division of labor is \textit{subject-based}. Among those who participate in tutoring, each caregiver takes responsibility for subjects they are familiar with and hand off those they are less confident in to others. For example, the mother tutors Chinese and English, the father tutors math and sports, and most grandmothers (if involved) are less literate than the parents so they \textit{``remind and supervise children to complete the homework, and only checks whether it's finished or not''.}
In contrast, in families with children in the lower primary grades (e.g., Grades 1-2), the division of labor is primarily \textit{availability-based}. Those who help with tutoring can handle any subject for younger grades, so when children have questions, \textit{``whoever is around helps; if one person is not available, another steps in.''}

\paragraph{\textbf{KI2: Limited collaboration between caregivers in family learning.}}
Although multiple family members may participate in daily family learning, coordination between them is often minimal. Tutoring tends to be parallel contribution, in which each caregiver is responsible for specific subjects or tasks, rather than engaging in joint problem-solving (e.g., co-tutoring in writing a composition).
Caregivers typically neither involve themselves in nor inquire about the tutoring conducted by others, and they rarely share updates on their own parts. 

One family attempted to coordinate tutoring through a WeChat group, however, due to the platform's high volume of messages, relevant information was easily overlooked, particularly by the elder family members.

\paragraph{\textbf{KI3: Imbalanced division of labor in family learning.}}
This imbalance in the division of labor is common in families that use an availability-based approach to tutoring. In such cases, one caregiver often becomes the primary tutor, carrying most of the workload until the other caregiver is available (e.g., one handles weekdays while the other takes weekends), or stepping in only during moments of high stress. As one participant explained: \textit{``I'm the one who mainly tutors her. If I start feeling overwhelmed, like when things get a bit chaotic, that's when the dad steps in, and we swap roles for a while.''}
Currently, families rely on subjective impressions rather than objective records of who tutored, when, and on what. As a result, contributions from the tutoring caregiver may be undervalued by less-involved family members. For example, one participant share: \textit{``The dad thinks the child's good grades are all self-taught, that they don't need any tutoring from caregivers to do so well.''}

\paragraph{\textbf{KI4: Parental guidance and restrictions on children's independent use of AI for learning.}} 
All participants expressed concerns about children's over-reliance on AI, noting that with such tools, \textit{``when children get a problem, they can just put it in, the answer comes out instantly, and they can simply copy it.''}
To address this concern, parents often act as gatekeepers, restricting or completely prohibiting children from using AI independently. Instead, they take an active, guiding role in how AI is incorporated into learning. We identified three main approaches.

First, parents guide the child's thinking and work through problems together before using AI. One participant, describing her method for tutoring composition, explained: \textit{``We don't let him throw the question straight into AI. We first talk about what he wants to write, maybe an event or a place, and together decide on the main idea. Only after the structure is clear might he use AI to enrich the content.''}
In mathematics, several participants suggested using AI to generate similar problems and prompting children to solve these variants as a way to reinforce learning. In this approach, AI functions as a supplementary resource rather than a shortcut to a finished answer.

Second, participants learn from AI and then teach children. This approach is common among caregivers whose own schooling was years ago, making it difficult to recall specific content or present it in age-appropriate way. By consulting AI, they can get age-appropriate solutions for their children's level, understand the solution process, then re-explain it to children. One parent suggested that, in addition to solutions, \textit{``If AI explained what the question is assessing and which concept it covers, I could grasp it more quickly and then teach it clearly.''}

Third, participants utilize AI as a review tool. After children complete the homework, several participants use AI to check answers, enabling them to monitor accuracy without re-solving every problem themselves. This approach provides oversight while saving time and effort.

\section{System Details}

Based on the identified key insights from the formative study, we derived four main Design Strategies (DS) to guide the design of system. 
\begin{itemize}
    \item \textbf{DS1: Task distribution.} Coordinate and enrich tutoring by dividing task list into subtasks matched to each caregiver's availability and expertise. 
    \item \textbf{DS2: Task coordination.} Facilitate collaboration and handover by allowing caregivers to share progress, context, and strategies, addressing the lack of inquiry or updates between them. 
    \item \textbf{DS3: Progress visualization.} Balance workload by tracking and visualizing each caregiver's time and contributions to ensure equitable responsibility. 
    \item \textbf{DS4: Tutoring support.} Support parent-led AI usage with features for guided problem-solving, generating similar questions, explaining in age-appropriate way, and reviewing answers before feedback. 
\end{itemize}

\subsection{System Overview} 
Informed by the design strategies, FamLearn (Parenting Partner for Daily Family Learning), an AI-empowered system, is designed and developed to foster recognition, shared accountability, and more engaged intra-family collaboration in daily learning. FamLearn is implemented as a mobile-responsive website with the frontend built on Vue.js and backend empowered by FastAPI. We utilized GPT-4o API from OpenAI to provide interaction features, design prompts to distribute tasks and provide more accurate tutoring support. For each family, participants are assigned a unique username and password to access FamLearn. FamLearn incorporates several features that aim to foster better coordination and collaboration among caregivers, while considering caregivers' availability and expertise to support their diverse needs in tutoring children's daily learning.

\begin{figure}[t]
  \centering
  \includegraphics[scale=0.04]{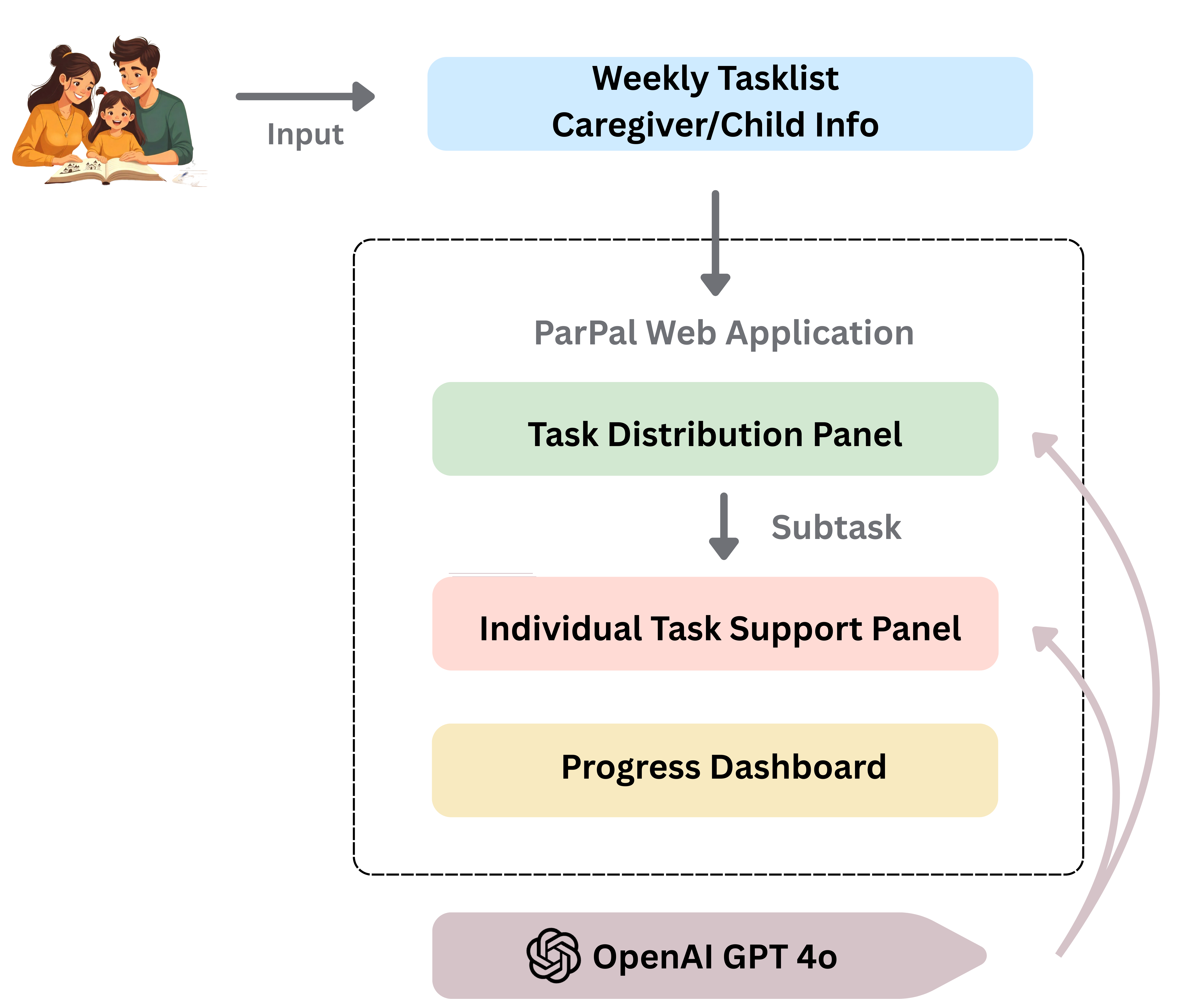}
  \caption{The system architecture of FamLearn. User input (weekly tasklist and caregiver/child information) flows into Task Distribution Panel, which assigns subtasks. The Individual Task Support Panel assists in completing them, while OpenAI GPT-4o supports the system by assisting task distribution and providing tutoring suggestions.}
  \Description{system architecture}
  \label{Figure1}
\end{figure}

As shown in Figure \ref{Figure1}, the following outlines the two main components of FamLearn: (1) the task distribution panel and dashboard to support prompt-based, personalized task distribution and collaboration among caregivers, visualized through a shared timesheet and displaying each caregiver's progress for the current week; (2) an individual task support panel offering specific and actionable instructions to complete each subtask, which we elaborate below.

\subsection{Task Distribution Panel}
\begin{figure}[t]
  \centering
  \includegraphics[scale=0.15]{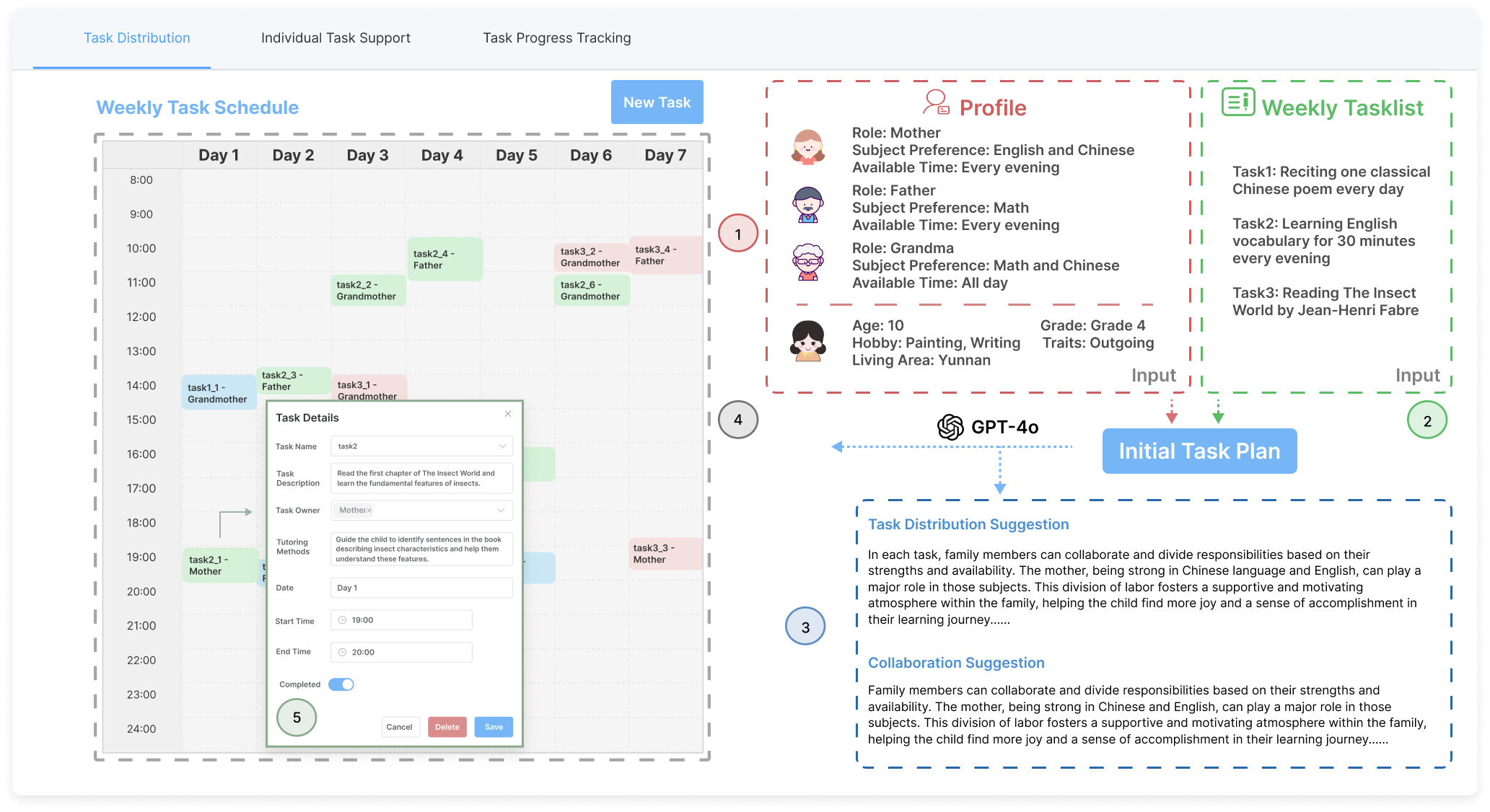}
  \caption{FamLearn's Task Distribution Panel to support task decomposition and allocation.}
  \Description{Task Distribution Panel}
  \label{Figure2}
\end{figure}

As identified in the design strategies \textbf{DS1} and \textbf{DS3}, task distribution panel (Figure~\ref{Figure2}) is designed to support collaboration and clear division of labor among family members for the child’s weekly learning. Meanwhile, this panel also visualizes the assigned tasks on a shared timesheet.

\subsubsection{Personalized task decomposition and allocation.}
Informed by the formative study, we identified that, in the context of family learning, division of labor and collaboration among different caregivers are closely related to their expertise and availability. To support this, FamLearn collects both caregivers' and children's information through a profile form completed by caregivers \textcircled{1}, which includes caregiver’s subject preferences, available tutoring time, and children's age, grade, hobbies, child’s traits and living area. Similarly, caregivers are also required to provide a task list for the upcoming week \textcircled{2}, describing what activities they expect their children to perform in the following week.

Once all profiles are configured, users can proceed to click on the \textit{Initial Task Plan} button to trigger the generation of personalized Weekly Task Schedule \textcircled{4}, along with task distribution and collaboration suggestions \textcircled{3}. FamLearn first decomposes the overall task list into multiple detailed and actionable subtasks when the task allocation feature is triggered. It then assigns these subtasks to different caregivers and generates task-specific tutoring suggestions. These suggestions provide a summary of a collaborative learning plan and specify how caregivers can work together to complete the assigned tasks. 

To achieve this, a \textit{multi-step prompting process} is designed to decompose, allocate subtasks, and support individualized tasks by incorporating profile information \textcircled{1} and weekly task list \textcircled{2} into prompts. This process consists of three main steps:
\begin{itemize}
    \item Step 1: Task decomposition. Firstly a prompt is designed to break down weekly task list into specific subtasks, focusing on generating specific subtask guidance that is both age-appropriate for children and aligned with caregivers' expertise and availability. Specifically, we adopt the role prompting method and define an expert persona with rich experience in the domain of family education, aiming to create more desirable outcome for task distribution in the context of family collaboration~\cite{wang2023rolellm}. 
    \item Step 2: Subtask scheduling. When obtaining specific and actionable subtasks, subtasks are allocated to caregivers and are scheduled for an appropriate day and time window according to their availability through another structured prompt, while maintaining logical task order.
    \item Step 3: Check for time conflicts. In the final step, the prompt is designed to help check time conflict problems, such as overlapping time slots. The final schedule plan is formatted in JSON and visualized in a shared weekly timesheet. In addition, a summary for weekly task schedule and collaboration suggestions are also generated to provide guidance to caregivers.
\end{itemize}

\subsubsection{Timesheet for sharing information.}
To address imbalanced division of labor in homework tutoring from formative study, a timesheet \textcircled{4} is designed to share task details. Meanwhile, when caregivers click on a subtask in the timesheet, a task card is displayed, showing the AI-generated task description, recommended tutoring method, task owner, expected start and end times, and task completion status \textcircled{5}. Due to variation in availability between working hours and personal time, FamLearn enables caregivers to adjust or hand over subtasks generated by AI to another caregiver.

\subsubsection{Dashboard for Task Progress Tracking and Summary}
\begin{figure}[t]
  \centering
  \includegraphics[scale=0.15]{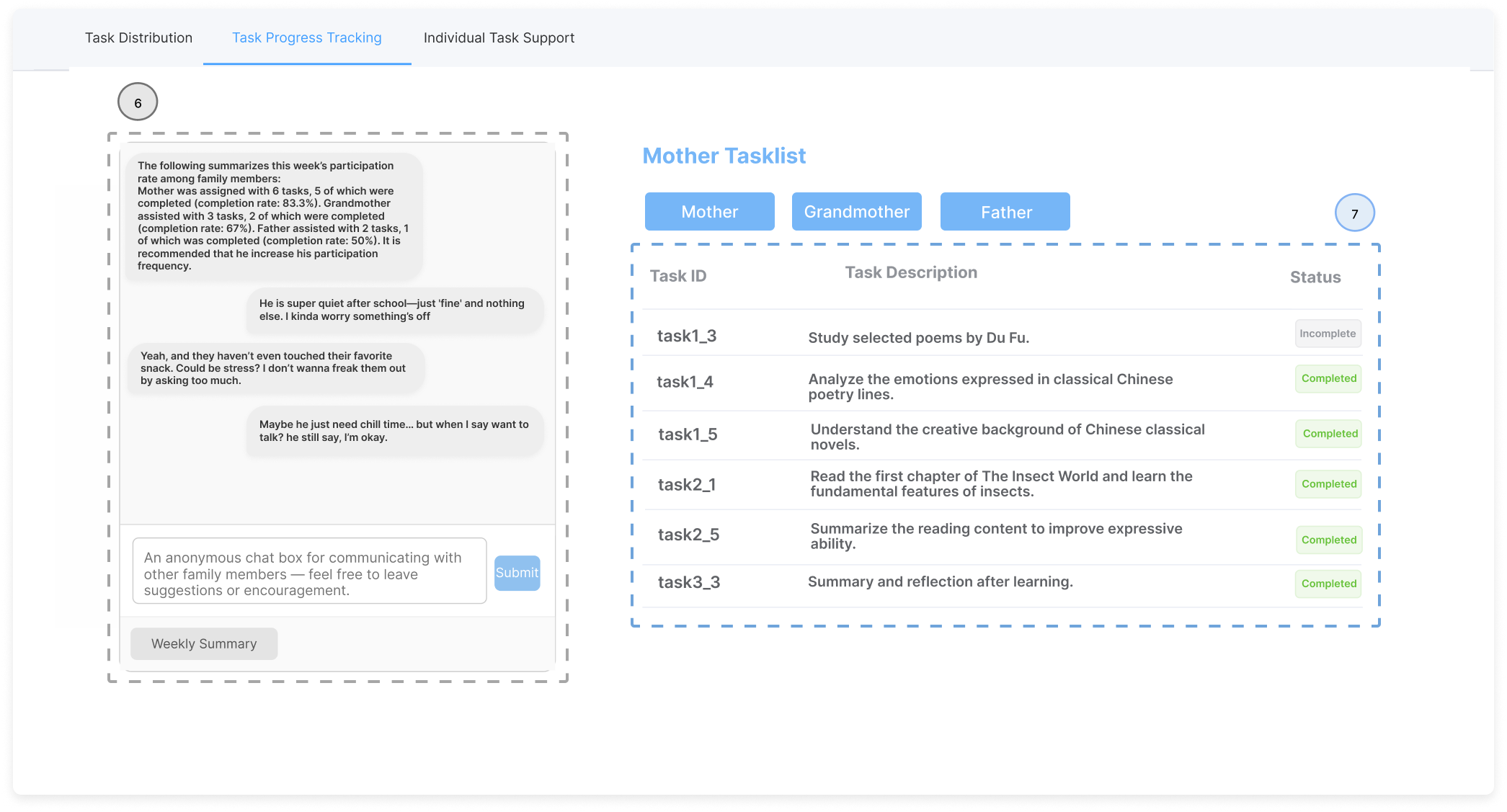}
  \caption{FamLearn's dashboard to support task progress tracking and summary.}
  \Description{dashboard }
  \label{Figure3}
\end{figure}

As shown in Figure \ref{Figure3} , we also create a dashboard to support caregivers to review each family member’s weekly progress and check the completion status of each task \textcircled{7}. This shared view helps caregivers monitor overall progress and stay informed about the contributions of other caregivers. 

\subsection{Individual Task Support Panel}
\begin{figure}[t]
  \centering
  \includegraphics[scale=0.15]{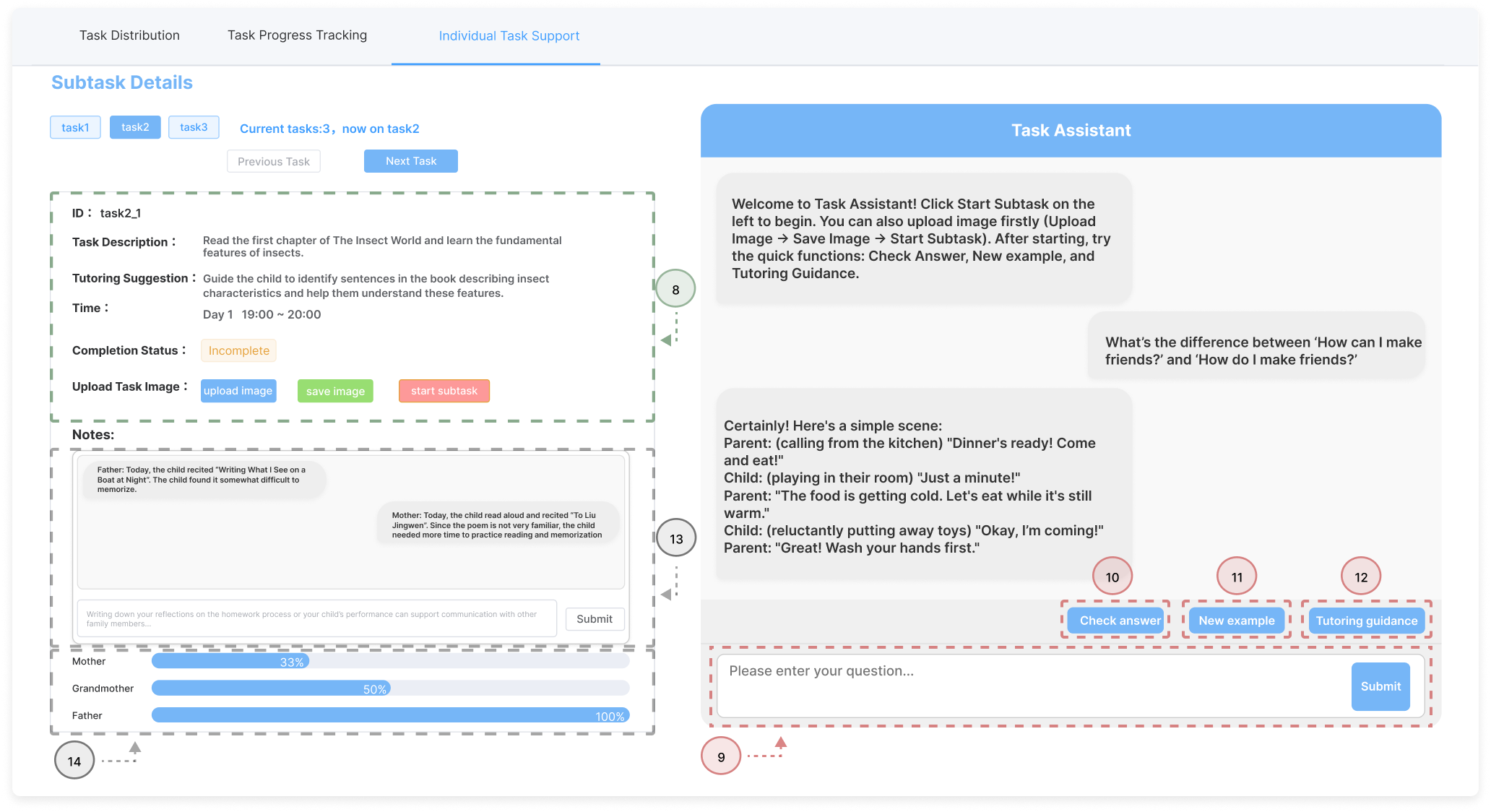}
  \caption{FamLearn's Individual Task Support Panel to support tutoring guidance.}
  \Description{Individual Task Support Panel}
  \label{Figure4}
\end{figure}

Another need identified from the formative study is to provide caregivers with direct support in everyday family learning, helping them overcome knowledge gaps and reducing the burden of tutoring (\textbf{DS4}). In this panel, an actionable task details \textcircled{8}, incorporating task description and tutoring suggestion, are presented to scaffold caregivers in tutoring children's learning. As shown in Figure~\ref{Figure4}, we also offer caregivers who share responsibility for the same task a space to exchange information, leave notes, or record progress \textcircled{5}.

\subsubsection{Interaction with tutoring assistance.}
When caregiver begins working on their assigned task, they can interact with a prompt-based chatbot powered by GPT-4o through text input \textcircled{9}. Caregivers could ask questions, seek clarification, or request supplementary materials to address problems encountered during tutoring process. 

\subsubsection{Task-specific support features.}
FamLearn contains three key features designed to alleviate the burden of homework tutoring and to support caregivers in overcoming knowledge gap. 
\begin{itemize}
    \item \textit{Answer checking}. Along with function that caregiver can upload pictures of the homework within the detailed task information card, the assigned caregiver can click on the answer-checking button for quickly verifying answer \textcircled{10}.
    
    \item \textit{New example}. The second feature, new example, provides caregivers with additional examples or tasks that share underlying concepts with the original questions \textcircled{11}. Informed by formative study, several participants suggested that AI could be utilized to generate similar questions for mathematics test, enabling them to assess whether the child has truly understood how to solve the problems.

    \item \textit{Tutoring guidance}.The third feature is designed to offer age-appropriate tutoring guidance, which helps caregivers explain problems using language and examples that are accessible and understandable to the child \textcircled{12}.
\end{itemize}

The features, answer checking and new example, are similar to existing homework assistance platforms, such as \textit{Zuoyebang} \cite{zuoyebang2025} and \textit{Xiaoyuan Souti} \cite{xiaoyuandouti2025}, which offer automated feedback upon uploading a photo and typically require paid subscriptions. However, tutoring guidance in FamLearn emphasizes age-appropriate explanations to the child’s grade level and cognitive stage and guides caregivers in how to explain problems in a structured way.

\section{User Study}

\subsection{Participants and Recruitment}
We recruited 11 families through prior contacts and snowball sampling. In total, 23 individuals
participated: 11 mothers, 10 fathers and 2 grandmothers, including 9 pairs of mothers and fathers, and 2 groups with both parents and grandmothers.
To meet our inclusion criteria, participants must be (1) have prior experience in tutoring children, and (2) have children currently attending elementary or junior high school. Participant demographics are summarized in Table~\ref{Demongraphic information}. Each household received 200 RMB ($\approx$30 USD) as compensation for their participation. This study is IRB-approved.

\begin{table}[]
\small
\caption{Participants' demographics.}
\label{Demongraphic information} 
\resizebox{\columnwidth}{!}{%
\begin{tabular}{p{1cm}p{1.5cm}p{1cm}p{1cm}p{2cm}p{2cm}p{3.2cm}p{0.5cm}p{0.6cm}p{0.8cm}p{0.5cm}}
\toprule
\textbf{Family ID} & \textbf{\shortstack{Caregiver ID$^{*}$}} & \textbf{Role} & \textbf{Age Group} & \textbf{Occupation}  & \textbf{Education Level} & \textbf{AI Learning Tools Used}& \textbf{Child ID} & \textbf{Grade} & \textbf{Gender} & \textbf{Age}

\\ \hline

\multirow{3}{*}{Family 1} & F1D & Father & 41-45 & Goverment Job & Undergraduate & \multirow{3}{*}{Zuoyebang, Doubao, Baidu AI} & \multirow{3}{*}{FK1} & \multirow{3}{*}{4} & \multirow{3}{*}{Female} & \multirow{3}{*}{10} \\ \cline{2-6}
& F1M & Mother & 36-40 & Goverment Job & Undergraduate &  &  &  &  &  \\ \cline{2-6}
& F1GM & Grandma & 61-65 & Retired & Not Disclosed &  &  &  &  &  \\ \hline
\multirow{2}{*}{Family 2} & \shortstack{F2M$^{*}$} & Mother & 36-40 & Teacher & Undergraduate  & \multirow{2}{*}{Zuoyebang} & \multirow{2}{*}{FK2} & \multirow{2}{*}{5} & \multirow{2}{*}{Female} & \multirow{2}{*}{11} \\ \cline{2-6}
& F2D & Father & 36-40 & Technical Staff & Undergraduate&  &  &  &  &  \\ \hline
\multirow{2}{*}{Family 3} & \shortstack{F3M$^{*}$} & Mother & 41-45 & Cost Consulting & Undergraduate & \multirow{2}{*}{Zuoyebang, Yuanfudao} & \multirow{2}{*}{FK3} & \multirow{2}{*}{4} & \multirow{2}{*}{Male} & \multirow{2}{*}{9} \\ \cline{2-6}
& F3D & Father & 41-45 & Landscaping & Junior College &  &  &  &  &  \\ \hline
\multirow{2}{*}{Family 4}& \shortstack{F4M$^{*}$} & Mother & 41-45 & Teacher & Undergraduate & \multirow{2}{*}{Zuoyebang, Yuanfudao} & \multirow{2}{*}{FK4} & \multirow{2}{*}{6} & \multirow{2}{*}{Male} & \multirow{2}{*}{12} \\ \cline{2-6}
& F4GM & Grandma & 70-74 & Retired & High School&  &  &  &  &  \\ \hline
\multirow{2}{*}{Family 5} & \shortstack{F5M$^{*}$} & Mother & 36-40 & Professor & PhD & \multirow{2}{*}{ChatGPT} & \multirow{2}{*}{FK5} & \multirow{2}{*}{3} & \multirow{2}{*}{Male} & \multirow{2}{*}{8} \\ \cline{2-6}
& F5D & Father & 36-40 & Engineer & PhD &   &  &  &  &  \\ \hline
\multirow{2}{*}{Family 6} & \shortstack{F6M$^{*}$} & Mother & 36-40 & Staff & Undergraduate & \multirow{2}{*}{Zuoyebang} & \multirow{2}{*}{FK6} & \multirow{2}{*}{5} & \multirow{2}{*}{Female} & \multirow{2}{*}{11} \\ \cline{2-6}
& F6D & Father & 36-40 & Staff & Undergraduate &  &  &  &  &  \\ \hline
\multirow{2}{*}{Family 7} & F7M & Mother & 41-45 & Teacher & Undergraduate & \multirow{2}{*}{Doubao, Ernie Bot} & \multirow{2}{*}{FK7} & \multirow{2}{*}{4} & \multirow{2}{*}{Female} & \multirow{2}{*}{10} \\ \cline{2-6}
& F7D & Father & 41-45 & Engineering & Undergraduate &  &  &  &  &  \\ \hline
\multirow{2}{*}{Family 8} & F8M & Mother & 36-40 & Teacher & Undergraduate & \multirow{2}{*}{Doubao} & \multirow{2}{*}{FK8} & \multirow{2}{*}{4} & \multirow{2}{*}{Male} & \multirow{2}{*}{10} \\ \cline{2-6}
& \shortstack{F8D$^{*}$} & Father & 36-40 & Engineering & Undergraduate &  &  &  &  \\ \hline
\multirow{2}{*}{Family 9} & \shortstack{F9M$^{*}$} & Mother & 36-40 & HR & Undergraduate & \multirow{2}{*}{Deepseek, Yuanfudao} & \multirow{2}{*}{FK9} & \multirow{2}{*}{6} & \multirow{2}{*}{Male} & \multirow{2}{*}{11} \\ \cline{2-6}
& F9D & Father & 41-45 & Lecturer & Graduate &  &  &  &  &  \\ \hline
\multirow{2}{*}{Family 10} & F10D & Father & 36-40 & Technical Staff & Undergraduate & \multirow{2}{*}{Zuoyebang, Yuanfudao} & \multirow{2}{*}{FK10} & \multirow{2}{*}{3} & \multirow{2}{*}{Male} & \multirow{2}{*}{9} \\ \cline{2-6}
& F10M & Mother & 36-40 & Telecommunications & Undergraduate &  &  &  &  &  \\ \hline
\multirow{2}{*}{Family 11} & F11D & Father & \multicolumn{1}{l}{41-45} & Security Officer & High School & \multirow{2}{*}{Baidu AI, Xiaoai classmate} & \multirow{2}{*}{FK11} & \multirow{2}{*}{5} & \multirow{2}{*}{Male} & \multirow{2}{*}{11} \\ \cline{2-6}
& \shortstack{F11M$^{*}$} & Mother & \multicolumn{1}{l}{36-40} & Teacher & Undergraduate &  &  &  &  &  \\ \bottomrule
\end{tabular}
}
{\raggedright \small* Primary caregiver, who takes the leading role in daily family learning. Without *, do not have a clearly identified primary caregiver.}
\end{table}

\subsection{Study Procedure}
\subsubsection{Pre-study onboarding.}
Before the formal study, each family was invited to join an online onboarding session. 
During the onboarding session, we introduced the study procedure and provided instructions on how to interact with our system. We explicitly explained the process and assisted participants in entering caregiver information, child information, and initiating task generation using the task list they had prepared in advance. Next, we described each panel in the system and demonstrated the functionalities designed to support parents in family learning. Participants were encouraged to adjust the AI-generated tutoring plans by adding, removing, or editing subtasks when the AI-assigned content did not align with their practical situation. We also encouraged them to explore how the Chatbot could help address challenges in daily family learning. 

\subsubsection{Field study.}
Each family was required to participate for one week, and six families voluntarily participated for two weeks. During the study, we checked in regularly to address any problems they encountered. Participants were encouraged to collaborate with other caregivers in tutoring their children’s family learning through \tool{} and to use it to support their daily tutoring practices. 

\subsubsection{Post-study interview.}
In the final phase of the study, we conducted post-study interviews with all families. Each interview lasted approximately 60 minutes and included all caregivers who had participated in the study. The purpose of the post-study interviews was to understand how caregivers used \tool{} to support their daily family learning. We asked questions about (1) their experiences interacting with \tool; (2) how \tool{} supported collaboration and communication among caregivers and assisted with their daily learning practices, with particular attention to task distribution, visualization through the timesheet, and the three tutoring functions; and (3) their perceptions of \tool’s contributions across several dimensions, including hidden caregiving labor, coordination of family opinions, visibility of caregiving roles, emotional support in parenting, children’s learning outcomes, and improvements in caregivers’ tutoring strategies. 

All interviews were originally conducted in Mandarin and audio recorded after consent. We analyzed the transcribed interviews using a thematic analysis approach \cite{braun2006using}. Two authors independently coded transcripts from the same five families, reviewed each other’s work, and discussed interpretations with the broader team to develop an initial codebook. As additional interviews were incorporated, the two authors refined and expanded the codebook, with regular team discussions ensuring that codes and interpretations remained consistent and grounded in the data. Final themes were consolidated through ongoing dialogue until the research team reached full consensus.

\subsection{Study Results}

\begin{figure}[t]
  \centering
  \includegraphics[scale=0.35]{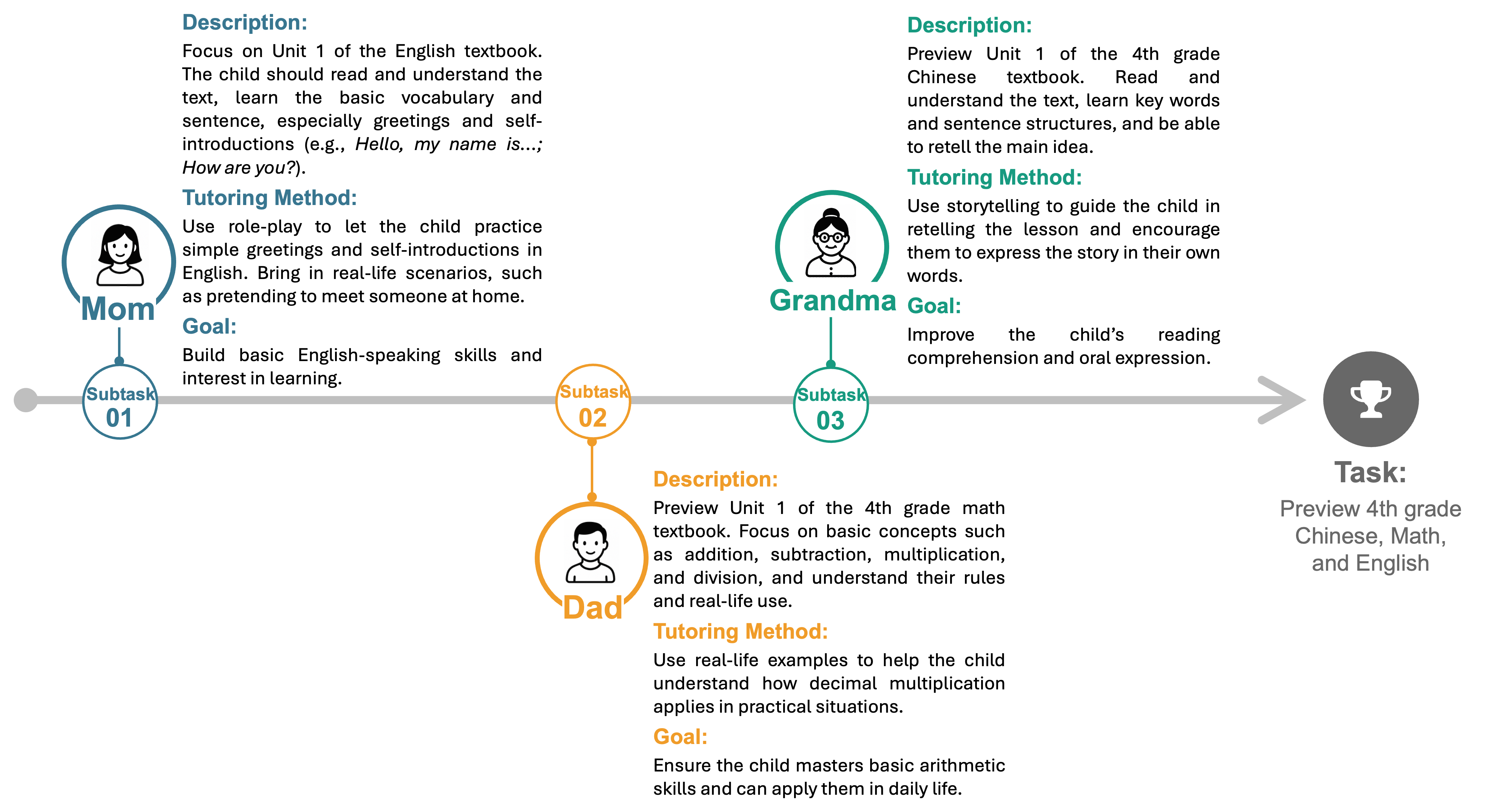}
  \caption{Example Tasks and Subtasks Completed by Family F1}
  \Description{Example Tasks and Subtasks Completed by Family F1}
  \label{Figure5}
\end{figure}

\begin{table}[!t]
\small
\caption{Task distribution and tutoring assistance across families and participants. 
\label{system usage} 
\textbf{Task Distribution} reports the number of completed tasks and executed subtasks. 
\textbf{Individual Task Support} records whether specific tutoring strategies were applied (new example, answer checking, tutoring guidance, interaction with Chatbot). Example Chatbot conversations are excerpted from log data.}
\resizebox{\textwidth}{!}{
\begin{tabular}{
>{\centering\arraybackslash}m{0.8cm}
>{\centering\arraybackslash}m{1.2cm}|
>{\centering\arraybackslash}m{1.2cm}
>{\centering\arraybackslash}m{1.8cm}|
>{\centering\arraybackslash}m{1.5cm}
>{\centering\arraybackslash}m{1.5cm}
>{\centering\arraybackslash}m{1.5cm}
>{\centering\arraybackslash}m{1.5cm}
>{\centering\arraybackslash}m{12cm}
}
\toprule
\multirow{2}{*}{\makecell{\textbf{Family}\\\textbf{ID}}}&
\multirow{2}{*}{\makecell{\textbf{Caregiver}\\\textbf{ID}}} &
\multicolumn{2}{c|}{\textbf{Task Distribution}} &
\multicolumn{5}{c}{\textbf{Individual Task Support}} \\
\cline{3-9} 
 &
   &
  \textbf{Tasks Completed (\#)} &
  \textbf{Subtasks Executed for Task Completion (\#)} &

  \textbf{Use of \textit{New Example \textcircled{10}}} &
  \textbf{Use of \textit{Answer Checking \textcircled{11}}} &
  \textbf{Use of \textit{Tutoring Guidance \textcircled{12}}} &
  \textbf{Use of \textit{Chatbot} (Default)} &
  \textbf{Example Conversation with Chatbot (from log data)} \\
\midrule

\multirow{3}{*}[-6ex]{F1} & \raisebox{-1.5ex}{F1D} & \multirow{3}{*}[-6ex]{6} & \raisebox{-1.5ex}6
& \raisebox{-1.5ex}Y & \raisebox{-1.5ex}Y & \raisebox{-1.5ex}Y & \raisebox{-1.5ex}Y & 
\multirow{3}{*}{
\makecell[{{p{\linewidth}}}]{
\textbf{F1M asked:} \textit{``How do you make a convex lens?''}\\
\textbf{Chatbot replied:} \textit{``Making a real convex lens needs professional optical tools, but you can simulate the effect with simple materials.
Materials: clear plastic sheet or glass, water, scissors, glue
Steps: (1) Prepare the plastic: Cut the plastic into a circle or oval; (2) Shape it: Bend the edges upward slightly to form a shallow bowl; (3) Add water: Fill the bowl-shaped plastic with water so the surface bulges outward, forming a convex curve; (4) Secure the shape: Use glue to attach the edges to a circular frame, locking in the shape. This simple model shows how a convex lens focuses light, though true lenses require special grinding and polishing.} 
}}\\[3.5ex] \cline{2-2}  \cline{5-8}
& F1M &  & 10  & Y & Y & Y & Y &\rule{0pt}{6.5ex} \\ \cline{2-2}  \cline{5-8}
& F1GM &  & 19 & Y & Y & Y & Y & \rule{0pt}{6.5ex} \\ 
\midrule

\multirow{2}{*}[-5ex]{F2} & \raisebox{-2ex}{F2D}  & \multirow{2}{*}[-5ex]{6} & \raisebox{-2ex}{23}
& \raisebox{-2ex}Y & \raisebox{-2ex}Y & \raisebox{-2ex}Y & \raisebox{-2ex}Y &
\multirow{2}{*}{
\makecell[{{p{\linewidth}}}]{
\textbf{F2D asked:} \textit{``What is the background behind ‘Guan Shu You Gan No.1’? How can we help the child understand the general meaning of the poem?''}\\
\textbf{Chatbot replied: }\textit{``Guan Shu You Gan No.1’ was written by the Song Dynasty poet Zhu Xi. It describes the sudden enlightenment that reading can bring. The poem's general meaning is: reading enlightens us, like parting clouds to reveal the blue sky. It reminds us of the importance of persistent learning. You can teach children that reading can bring wisdom and new perspectives.''}\\ 
}}\\[6ex] \cline{2-2}  \cline{5-8}

& F2M & & 31 & Y & Y & Y & N &\rule{0pt}{8ex} \\
\midrule

\multirow{3}{*}[-5ex]{F3} & \raisebox{-2ex}{F3D} &  \multirow{3}{*}[-5ex]{4} & \raisebox{-2ex}{11}
& \raisebox{-2ex}Y & \raisebox{-2ex}Y & \raisebox{-2ex}Y & \raisebox{-2ex}Y &
\multirow{3}{*}{
\makecell[{{p{\linewidth}}}]{
\textbf{F3M asked:} \textit{``My child is reading Hans Christian Andersen's Fairy Tales. How can I encourage him to read?''}\\
\textbf{Chatbot replied: }\textit{``You can try the following methods. (1) Story Time: Set aside a fixed time each day to read together to create a warm atmosphere. (2) Role Play: Role-Play with your child to increase interactivity. (3) Discussion: After reading, discuss the plot and characters and allow your child to express his own opinions. (4) Creative Activities: Encourage your child to draw his favorite scenes or characters. (5) Rewards: Use small rewards to motivate consistent reading.''}\\ 
}}\\ [6ex]\cline{2-2}  \cline{5-8}
& F3M & & 7 & Y & Y & Y & N &\rule{0pt}{8ex} \\
\midrule

\multirow{3}{*}[-1ex]{F4} & \raisebox{-1ex}{F4GM} & \raisebox{-4ex}3 & \raisebox{-1ex}5
& \raisebox{-1ex}Y & \raisebox{-1ex}Y & \raisebox{-1ex}Y & \raisebox{-1ex}Y &
\multirow{3}{*}{
\makecell[{{p{\linewidth}}}]{
\textbf{F4GM asked:} \textit{``Playing guitar for fun.''}\\
\textbf{Chatbot replied: }\textit{``Playing guitar is a great hobby that helps children relax and improve their musical skills. Consider adding a fixed practice time to your child's schedule so they can enjoy practicing guitar while enjoying basketball class. This helps balance their interests.''}\\ 
}}\\ [2.5ex]\cline{2-2}  \cline{5-8}
& F4M & & 4 & N & N & Y & Y & \rule{0pt}{5ex}\\
\midrule

\multirow{3}{*}[-8ex]{F5} & \raisebox{-4ex}{F5D} & \raisebox{-11ex}1 & \raisebox{-4ex}2
& \raisebox{-4ex}N & \raisebox{-4ex}N & \raisebox{-4ex}N & \raisebox{-4ex}N &
\multirow{3}{*}{
\makecell[{{p{\linewidth}}}]{
\textbf{F5M asked:} \textit{``Come up with a quick little game to help with English reading for saving time while tutoring''}\\
\textbf{Chatbot replied: }\textit{``Sure! Based on family information and the child's interests, I've designed a mini-game that can be played during tutoring and improve child's English reading skills in a relaxed and enjoyable atmosphere. 
Game Name: Lego Story Chain 
Objective: Through storytelling, strengthen children's English reading and speaking skills while sparking imagination. 
Materials Required: Some Lego bricks, A simple English storybook (suitable for third graders), A whiteboard and markers (optional, for recording story plots) 
Steps: 1. Choose a Story: Pick a storybook for third graders on a topic they enjoy, like Pokémon
\ldots
Thursday Night: Dad can join in and help the child use mathematical logic to reason about the storyline. We hope this game brings joy and helps improve your child’s English. Let me know if you'd like any changes.''}\\ 
}}\\ [10ex]\cline{2-2}  \cline{5-8}
& F5M & & 5 & Y & N & Y & Y & \rule{0pt}{13ex} \\
\midrule

\multirow{3}{*}[-8ex]{F6} & \raisebox{-4ex}{F6D} & \raisebox{-10ex}4 & \raisebox{-4ex}4
& \raisebox{-4ex}N & \raisebox{-4ex}N &\raisebox{-4ex} N & \raisebox{-4ex}Y &
\multirow{3}{*}{
\makecell[{{p{\linewidth}}}]{
\textbf{F6D asked:} \textit{``With the start of sixth grade approaching, what preparations are recommended for family members?''}\\
\textbf{Chatbot replied: }\textit{``As sixth grade approaches, there are several things you can do to help your child prepare in academics, daily routines, and mindset. Here are some suggestions:
Academic Preparation: 1. Review and Preview: Mom can use her strengths in Chinese, math, and English to help the child review what was learned in 5th grade and preview 6th-grade materials. This can be done in the evenings, focusing on one subject at a time to avoid overwhelming the child}\ldots\textit{Feel free to share feedback so I can adjust the plans to fit your child and family''}\\ 
}}\\ [9ex]\cline{2-2}  \cline{5-8}

& F6M & & 14 & Y & Y & Y & Y & \rule{0pt}{11ex} \\
\midrule

\multirow{3}{*}{F7} & F7D & \raisebox{-3ex}1 & 1
& N & N & Y & N &
\multirow{3}{*}{
\makecell[{{p{\linewidth}}}]{
\textbf{F7D asked:} \textit{``Please summarize the current assignments and provide the assignment titles.''}\\
\textbf{Chatbot replied: }\textit{``Guide the child to choose the text or passage for morning recitation. Which text or passage would you like to recite this morning?''}\\ 
}}\\ [1.5ex]\cline{2-2}  \cline{5-8}
& F7M & & 1 & N & N & N & N & \rule{0pt}{3.5ex}\\
\midrule

\multirow{3}{*}[-5ex]{F8} & \raisebox{-3ex}{F8D}&  \raisebox{-7ex}1 & \raisebox{-3ex}2
& \raisebox{-3ex}N & \raisebox{-3ex}N & \raisebox{-3ex}N & \raisebox{-3ex}Y &
\multirow{3}{*}{
\makecell[{{p{\linewidth}}}]{
\textbf{F8M asked:} \textit{``Dad is going to buy beer. Each winning bottle cap can be used to purchase a beer for 1 yuan. Normally, a bottle of beer costs 6 yuan. Dad has 3 winning bottle caps and wants to buy 6 bottles in total. How much will it cost?''}\\
\textbf{Chatbot replied: }\textit{``Dad has 3 winning bottle caps, which he can use to buy 3 bottles of beer at the discounted price of 1 yuan each. The remaining 3 bottles must be purchased at the regular price of 6 yuan each. So the total cost is: 3 bottles × 1 yuan + 3 bottles × 6 yuan = 3 yuan + 18 yuan = 21 yuan.''}\\ 
}}
\\ [6ex]\cline{2-2} \cline{5-8}
& F8M &  & 2& N & Y & Y & Y &  \rule{0pt}{8ex}\\
\midrule

\multirow{2}{*}[-4ex]{F9} & \raisebox{-2ex}{F9D} & \raisebox{-5ex}1 & \raisebox{-2ex}1
& \raisebox{-2ex}N & \raisebox{-2ex}Y & \raisebox{-2ex}N & \raisebox{-2ex}Y & 
\multirow{3}{*}{
\makecell[{{p{\linewidth}}}]{
\textbf{F9M asked:} \textit{``Could you give me some English phrases that sixth graders should know?''}\\
\textbf{Chatbot replied: }\textit{``Here are some examples: (1) make a decision, (2) have a good time, (3) in front of, (4) on time, (5) take part in, (6) pay attention to, (7) be interested in, (8) get along with, (9) look forward to, (10) turn on / turn off. These phrases help enhance children's ability to express themselves in English.''}\\ 
}}\\ [3.5ex]\cline{2-2}  \cline{5-8}
& F9M & & 2 & N & Y & N & Y &\rule{0pt}{5.5ex} \\
\midrule

\multirow{2}{*}[-2ex]{F10} & F10D & \raisebox{-2ex}1 & 1
& N & Y & N & Y &
\multirow{3}{*}{
\makecell[{{p{\linewidth}}}]{
\textbf{F10D asked:} \textit{``Give me some math Olympiad problems for a 3rd grader.''}\\
\textbf{Chatbot replied: }\textit{``(1) Xiaoming has 15 candies and shares with 5 friends \ldots }\textit{These problems can help children practice basic multiplication and division concepts.''}\\ 
}}
\\ [1.5ex]\cline{2-2}  \cline{5-8}
& F10M & & 2 & N & Y & N & N & \rule{0pt}{4ex} \\
\midrule

\multirow{2}{*}[-6ex]{F11} & \raisebox{-3ex}{F11D} & \raisebox{-6ex}1 & \raisebox{-3ex}4
& \raisebox{-3ex}N & \raisebox{-3ex}N & \raisebox{-3ex}N & \raisebox{-3ex}N &
\multirow{3}{*}{
\makecell[{{p{\linewidth}}}]{
\textbf{F11M asked:} \textit{``Could you come up with some new examples based on current questions?''}\\
\textbf{Chatbot replied: }\textit{``(1) Review this week’s Chinese text and answer these questions: What is the main theme? What message is the author trying to convey? (2) Think about another Chinese text you've read recently and answer these: What’s the setting of the story? What feeling do you think the author wanted to express through it? (3) Revisit this week’s Chinese text again and answer the following: What do the main characters experience? What idea or emotion is the author showing through their journey?'}\\ 
}}\\ [6ex]\cline{2-2}  \cline{5-8}
& F11M & & 4 & Y & Y & N & N & \rule{0pt}{8ex}\\
\bottomrule
\end{tabular}}
\Description[Task distribution and tutoring assistance]{This table shows, for each family (F1--F10), the number of tasks and subtasks completed, illustrative task/subtask descriptions, tutoring assistance strategies (new example, answer checking, tutoring guidance, interaction with chatbot), and representative chatbot conversation excerpts.}
\label{tab:family-tutoring}
\end{table}

\subsubsection{System Usage Analysis}
\label{finding_usage}

In total, all families collaboratively completed 29 major tasks and 160 individualized subtasks (an average of 3 major tasks and 14 subtasks per family). With all caregivers participating in each major task, the amount of completed subtasks was evenly distributed across caregivers in each family, and each caregiver's participation rate reached as high as 77\%. While all families were engaged with Chatbot, a total of 239 conversation rounds with Chatbot were logged, an average of 21 interactions per family. Specifically, among the three designed features, 7 families used the \textit{New Example feature \textcircled{10}}, 9 families used \textit{Answer Checking \textcircled{11}} and \textit{Tutoring Guidance \textcircled{12}}. Additionally, those interactions with the Chatbot covered a wide range of topics (e.g., English and mathematics learning, classical Chinese poetry learning , idiom explanation), and Table~\ref{system usage} presents an overview of each family's engagement with FamLearn and examples of their conversation logs. 

As shown in Figure~\ref{Figure5}, Family 1, consisting of a father, mother, and grandmother, collaboratively completed the task \textit{Preview 4th grade Chinese, Math, and English}. After task decomposition and allocation, the mother took primary responsibility for the English section, the father led the math section, and the grandmother handled the Chinese section. Meanwhile, each subtask included a description, tailored tutoring methods, and specific learning goals to support and guide caregivers in tutoring their children.

\subsubsection{Caregiver-Child Interaction: From Abstract Goals to Actionable Learning Pathways} 
\label{finding1}

This section highlights how \tool{} enhanced caregiver-child interaction. 
Participants indicated that the system turned abstract learning goals into concrete, age-appropriate steps. This supported the cultivation of multi-dimensional competence beyond homework, eased caregivers' cognitive burden in tutoring planning, and fostered children's engagement in family learning.

\paragraph{\textbf{AI-Supported Cultivation of Multi-dimensional Competence Beyond Homework}}

Five families (F1, F2, F3, F4, F10) noted that \tool{} goes beyond assisting with right-or-wrong homework questions, extending to the cultivation of children's broader competencies, including life skills, cultural literacy, and sports. It breaks down these abstract goals into concrete, age-appropriate steps. 

For instance, F3 utilized \tool{} to foster the child's independent living abilities. The tool suggested practicing safe knife use by cutting fruit together with parents. Over time, the child progressed from needing assistance to acting independently, as F3M recalled, \textit{``Last week he peeled some fruit all by himself, and he felt really proud of it, and shared the peeled fruit with us, that made him really happy.''}
F3D further underscored the importance of this multi-dimensional cultivation beyond academics, reflecting that \textit{``Kids today are under heavy academic pressure and spend most of their time studying, so they end up lacking basic life skills... That's why I think this tool is valuable. It encourages us, as parents, to set a goal every so often... to teach our child a specific skill and then follow through on it.''}

Similarly, Families F2 and F10 leveraged \tool{} to deepen the child's cultural literacy. They inquired about the background, meaning, and imagery of classical Chinese poems as a way to complement \textit{``the dull and boring process of a child memorizing poems word by word''} (F2M). 

In the domain of sports, Family F4 highlighted how \tool{} offered clear, practical guidance for basketball practice, such as warm-up stretches, post-practice recovery routines, and strategies for injury prevention.

\paragraph{\textbf{AI Mediation of Caregiver's Cognitive Burden in Tutoring Planning}}

Four families (F1, F3, F5, F9) appreciated that \tool{} alleviated the cognitive load of tutoring by structuring learning plans around caregivers’ availability and expertise. Rather than spending time and effort deciding \textit{``what to teach today''}, caregivers could simply follow the system's decomposed tasks and pre-set schedules.
As F3D reflected, his planning burden was reduced through the tool’s organized, progressive task design: 
\textit{``It's been quite helpful... with this tool our plans have become more structured, and the tutoring feels more targeted and orderly. Following the sequence step by step makes a real difference for the child’s learning. It also encourages us as parents to take tutoring more seriously and avoid procrastination.''}

Beyond structuring content, \tool{} also mediated caregivers' uncertainty about \textit{``how to teach''} by recommending resources tailored to children's grade level and knowledge base. By reducing the trial-and-error of selecting suitable materials, it further lightened caregivers' cognitive burden in tutoring. 
For example, when a task required English reading (what to teach), \tool{} suggested grade-appropriate books and films. As noted by F7M, \textit{``The tutoring guidance is really detailed. The task doesn't just say `do some reading,' it also tells you which book to pick and how to interact with the child, like discussing the storyline. That directly solves the problem of not knowing how to teach.''}

\paragraph{\textbf{AI-driven Task Structuring for Engaged Learning}}

By using homework as a starting point, \tool{} restructured tasks to guide learning beyond merely outcome-driven goals (e.g., getting the homework done). Instead, it emphasized process-centered learning that foregrounds understanding, reasoning, and reflection. This encouraged caregivers from six families (F1, F3, F6, F7, F8, F10) to engage more deeply with children in learning.

By providing age-appropriate, easy-to-understand solutions, \tool{} enabled caregivers to guide children more effectively and foster engagement in problem-solving. This contrasted with \textit{``calling and asking teachers for solutions previously''} (F3M) when without \tool. 
As F4GM noted, \textit{``the tool is helpful to help me and the child understand and solve a problem, rather than just check and give the answers.''} 
Similarly, F1M reflected that AI-suggested explanations not only improved comprehension but also reduced stress: \textit{``After consulting the AI, I tried a different approach, and [the child] understood it. In those moments, I really felt the tool was helpful. It calmed me down, because I saw it wasn’t that he didn’t understand, but that my explanation wasn’t effective.''}

Building on this scaffolding, children also began moving from \textit{``just completing homework''} to \textit{``actively understanding and analyzing.''} For example, F7M described how reading instruction shifted from simply finishing a text to deeper engagement: while reading \textit{Charlotte’s Web}, her child started analyzing character motivations and raising new questions. Such moments illustrate how structured guidance made tutoring a more reflective and interactive learning process.

\subsubsection{Caregiver-Caregiver Interaction: From Individual Efforts to Collaborative Family Practices} 

\label{finding2}
This section explores how \tool{} contributed to caregiver-caregiver interaction, shifting from individual efforts to more collaborative family practices. 
As F3M reflected, \textit{``it makes me see and be seen,''} pointing to how the system surfaced hidden efforts, supported task allocation and shared goals, and reduced tension and stress between family members.

\paragraph{\textbf{Increased Awareness of Hidden Efforts in Family Work}}

By documenting caregivers' completed tutoring tasks, six families (F1, F2, F3, F7, F9, F11) reported that the system made each parent’s contributions to supporting their child’s learning more visible. This visibility fostered greater recognition of one another's hidden efforts, which were often overlooked previously. 

For instance, F3D reflected that his wife's sustained involvement in tutoring was indispensable, even though it often went unnoticed in daily life. He remarked that without her consistent efforts, \textit{``the household would easily fall into disorder.''} He added that such tasks are always present, but in daily life they are \textit{``unintentionally avoided''} and only become visible when parents actively collaborate. 

F3M, in turn, admitted that she sometimes undervalued her husband's participation, especially when his approach to tutoring did not match her own expectations. After seeing his completed tasks recorded in the system, however, she recognized that \textit{``he was actually doing a great deal.''} This reminded her to acknowledge his contributions more fairly.

\paragraph{\textbf{Enhanced Parental Collaboration through Task allocation and Shared Goals}}

From the perspective of primary caregivers, the system encouraged greater involvement from previously less active caregivers. In the past, these family members often lacked clarity about how they could contribute. With task allocation and clear guidance, however, they could simply follow their assigned responsibilities, which fostered a stronger sense of accountability and participation. 
F6M revealed that the predefined tasks provided concrete entry points for her husband's participation: \textit{``when the system assigns tasks to the father, it gives him ideas and suggestions for how he can contribute.''}

F2M observed that while her husband had occasionally helped before, \textit{``after tasks were allocated, he felt a stronger sense of responsibility and became more willing to cooperate.''} She also highlighted that the structured tasks encouraged them to accompany their child together, which in turn strengthened family bonds: \textit{``the biggest change was that we followed the assigned tasks with our child, and it even improved our family activities, giving us more time to do things together.''}

\paragraph{\textbf{Emotional Support in and beyond Tutoring}} 

Several families (F1, F3, F6, F7) highlighted how \tool{} contributed to the emotional well-being of caregivers themselves, helping them manage stress and sustain cooperation at home. In F3M's case, tasks from \tool{} served as a reminder to regulate emotions after exhausted work: \textit{``When we return from work, sometimes we are tired and impatient. But through this tool, it tells you how to approach tasks, and then I consciously adjust my emotions, reminding myself not to bring negative feelings to the child or the family.''} 
\tool{} provided concrete references on what to do and how to do it, which helped caregivers calm down themselves rather than coming home feeling at a loss, carrying negative emotions, and struggling to communicate well with other family members and children.

Other families emphasized that, based on each caregiver's expertise and availability, \tool's task allocation and task-based guidance reduced disputes and made cooperation smoother. F7M explained that once responsibilities were divided, \textit{``each person knew what they should do.''} For example, when choosing reading materials, disagreements about whether to pick more difficult or easier books were resolved by the system's recommendations like \textit{The Little Prince}. Both parents readily agreed these were age-appropriate, reaching consensus without conflict. Beyond conflict reduction, F7M also noted that the suggested tutoring method and task progress tracking provided her with ongoing support, which eased her parenting-related anxiety.

\subsubsection{Constraints of AI in Informal and Varied Learning Contexts}
\label{finding3}
This section examines the constraints of \tool{} in informal and varied learning contexts, where occasional learning practices could not always be fully captured or supported. Participants mentioned challenges in reflecting caregiving efforts beyond quantifiable metrics, and expectations for more child-centered systems with parental monitoring features.

\paragraph{\textbf{Challenges in Supporting Incidental Learning Opportunities}}
\label{findings_incidental_learning}

Participants highlighted that while \tool{} offers structured support for planned learning tasks, it falls short in capturing and scaffolding unplanned, incidental opportunities for teaching that arise in everyday activities. These are moments where parents naturally integrate learning into flexible, non-predefined daily routines, such as during grocery shopping, sports, or casual family outings.

F5M illustrated this limitation by contrasting the tool's task-based design with spontaneous learning opportunities. She provided an example of informal math learning related to calculating payment and change at the checkout counter: \textit{``Some things can’t really be captured, because these tasks all need to be manually checked off. But when we go shopping, I might let him figure out how much money we need to pay, or if I pay cash, how much change I should get back. These activities involve real calculation, and they get him thinking, but they can't be put into the task list.''}

Similarly, H11M described how incidental learning could emerge during moments of daily care routines. For instance, during bath time, the father would use the opportunity to talk with his child about recent changes, personal interests, or broader life topics. These conversations were less about structured teaching, yet they still created meaningful spaces for child's learning, reflection and growth.
These example shows how spontaneous, real-world experiences often give children chances to think and practice skills, but \tool{} cannot easily notice or support these moments.

\paragraph{\textbf{Difficulties in Capturing Caregiving Efforts Beyond Quantifiable Metrics}}

Participants noted that while \tool{} currently provides metrics such as time spent, completion rate, and frequency of tutoring activities, these quantifiable measures only capture a narrow proportion of caregiving efforts. They pointed out that the caregiving responsibilities are shaped by context, shifting roles, and everyday interactions that cannot be reduced to metrics.

F9M cautioned that breaking down contributions purely through quantifiable tutoring data could unintentionally reinforce gender norms. In some households, mothers often spend more time with children due to work arrangements, the system's numbers may appear to favor them, which she warned could \textit{``easily turn into a gender confrontation.''} She further noted that a caregiver's tutoring involvement also fluctuates with other responsibilities and pressures: \textit{``Maybe the father has his own difficulties this week, work-related or otherwise. He has his own stress, and needs to adjust himself.''} In such situations, lower tutoring input should not be mistaken as lower commitment.

F6M also found that these metrics can only offer a rough snapshot, not a complete picture of family participation. This gap becomes especially salient when incidental learning opportunities (Section~\ref{findings_incidental_learning}) are taken up by non-primary caregivers. As she explained, \textit{``When the grandfather picks up the child and talks through a question, or when the father uses something on the street to explain a life lesson, that is also education, but these temporary situations cannot be recorded.''} Such contributions remain invisible in the system's metrics. If only primary caregivers' formal tutoring tasks are counted, the overall picture could be skewed and undervalue the less frequent but also meaningful involvement of other family members.

\paragraph{\textbf{Expectations of Child-Centered Systems with Parental Monitoring Features}}

The formative study (Section~\ref{formative_insights}) revealed that caregivers often restrict or even prohibit children from using AI independently for family learning. Based on this insight, we developed \tool, an AI system designed primarily to support caregivers in children’s learning, rather than being directly used by children. However, after the field study, participants indicated that certain functions could in fact be child-centered with parental monitoring function.

Caregivers envisioned AI as a means of scaffolding children’s autonomy while still keeping them informed. As F10D put it, \textit{``The design should focus on fostering children’s interest in completing tasks, rather than relying on parental tutoring. Since parents are busy, we hope to track children's progress during work hours and remind them remotely to complete tasks.''}

F9M further elaborated that the real challenge lies not in the usefulness of AI tools but in children's inefficiency and motivation. In her words, problems with specific assignments are usually solvable, but ``human factors'' like \textit{``children being distracted''} consistently hinder learning. She emphasized that this issue was also encountered by other families: 
\textit{``The real challenge... is the child’s own inefficiency. When we talk with other parents, the conversation is rarely about how to solve a particular problem or which questions the child can’t handle. Those aren’t real issues. Anything directly related to learning usually has a solution. What is hard to resolve are the human factors.''}

To bridge this gap, F1M described how some functions of the system could be directly accessible to children, with parents monitoring child-AI interactions. For instance, parents wanted visibility into what children asked the AI and how the AI responded, and allowed children limited flexibility to adjust their tasks. A child might, for example, reduce the number of tasks when feeling unwell or overloaded. Parents, upon review, could then evaluate whether to recognize the completion of this adjusted goal. This design is child-centered, which grants children a degree of autonomy while ensuring parents remain in a supervisory role. 
\section{Discussion}

\subsection{From Single Parent-Child Dyadic to Family-Centered Learning to AI-supported Family Learning}
\subsubsection{A Shift from Single Parent-Child Dyadic to Family-Centered learning.}
In the context of family learning, collaboration should involve both children and parents, while allowing each family member to contribute to the child’s growth. 
In our formative study, caregivers initially expressed concerns about children using AI learning tools independently, due to worries about over-reliance and distraction on AI tools. However, findings from our field study (Section~\ref{finding3}) revealed a shift in attitude: caregivers became more willing to involve children in the use of the AI learning system. This change may stem from differences between existing learning tools and FamLearn. Most current learning apps are designed for general purposes and are not tailored to specific learning contexts, which raises concerns among caregivers about distraction and overuse. However, FamLearn is designed as a task-specific learning tool that supports structured, family-centered collaboration. Its concrete, guided approach helps caregivers feel more in control of the learning process and alleviates concerns about children’s overdependence on AI.

By reducing parents' concerns about the usage of AI tools in learning, different mediation strategies could be adopted, such as restrictive mediation, active mediation, and co-using~\cite{yu2021parental}. While such caregivers' mediation strategies in varying level could help reduce the concerns about children's usage of AI, they also offload the education burden from caregivers by granting children appropriate agency. Additionally, prior work has highlighted the need to turn personal informatics into family informatics~\cite{pina2017personal}. Our findings extend this perspective by emphasizing that safety is central to enabling parents to include children in collaborative educational practices. Accordingly, we suggest that parental controls, alongside mediation strategies, should be integrated into the design of AI learning tools. Such designs can foster caregiver-child collaboration, balancing children’s safety with reducing caregivers’ educational burden.

\subsubsection{AI Roles in Family-Centered Learning.}
As AI technologies become increasingly involved in the family context, understanding what roles AI should take in daily family learning is important to support effective and balanced collaboration among caregivers. Instead of replacing human involvement, AI systems should engage adaptively, and respond to each family’s unique dynamics and needs. Based on caregiver's availability, expertise and involvement, AI has the potential to play with three roles \cite{huang2025toward}. First, as the role of \textit{Advisor}, AI could offer short, age-appropriate explanation and learning-related resources, while still letting caregivers remain in control. 
Second, when tasks involve multiple steps or need coordination between caregivers, AI could act as a \textit{Co-Pilot}, which tracks progress, identifies what to do next, and makes it easier for one caregiver to hand over the task to another. 
Third, in some situations, AI could take the role of a \textit{Guardian} by avoiding giving direct answers too early and instead encouraging the child to try first, or by alerting caregivers when the child’s focus dropped. Specifically, in the role of Guardian, AI waits to give answers until the child has made an effort or reminds the caregiver when the child seems distracted. These dynamically shift between roles potentially help caregivers better understand how and when to rely on AI in collaborative family learning, avoiding over-reliance that may hinder child learning or under-utilization.

Additionally, a uniform support strategy may not be effective, since the caregivers have different expertise, fluctuating availability, and stress levels. To meet these diverse family learning contexts, the AI systems should be more adaptive and provide timely tutoring support. As shown in Section~\ref{finding1}. We also observe that FamLearn supports shifting caregivers’ focus from simply “getting the homework done” to helping children better understand and reason. AI systems could adjust the level and format of assistance based on real-time workload or preferences, offering shorter and more focused hints during busy evenings, while providing more reasoning support during planned study sessions.

\subsection{Awareness for United Collaboration}

Understanding how to effectively represent family members’ participation in collaborative caregiving is essential for supporting equitable involvement. While it is critical to know what metrics are appropriate for capturing family members' participation in family activities, current metrics tend to emphasize tutoring-related participation rather than providing a comprehensive measure of caregiving efforts. Specifically, some incidental learning opportunities and invisible caregiving contributions, such as moments where caregivers naturally integrate learning opportunities into daily routines like grocery shopping, are difficult to capture. Prior work has shown that if the visualization of efforts does not reflect the contributors' perceived efforts, they withdraw from the act of tracking the effort differences based on the visualization~\cite{dimicco2004influencing}. Moreover, conflicts may arise with single-dimensional metrics, as collaborators might only evaluate efforts based on the numbers~\cite{kelly2016can}. While previous studies have suggested that married couples consider each other's mental state when collaborating for the family~\cite{lin2024coparenting, lin2025parental}, our findings further suggest that caregiving responsibilities must also take into account context, shifting roles, and everyday pressures, such as workload and stress. Not to mention that incidental efforts are difficult to record (Section~\ref{finding3}). Future research could explore how to incorporate these factors and investigate approaches that help family members track incidental efforts.

In addition, different visualizations of efforts can lead to different interpretations. Participants were concerned that the current visualization reinforces gender norms, as fathers and mothers may compare their contributions based only on their individual inputs. This may stem from the fact that the efforts are portrayed as individual rather than collective.
Prior work has found that visualizing collaborators' efforts as personal representations is less favorable than visualizing accumulative efforts or a mix of personal representations and collective efforts~\cite{tausch2016comparison}. Similarly, research has shown that presenting efforts from the perspective of shared experience can foster interpersonal relationships and facilitate collaboration~\cite{lin2025parental}. Therefore, to mitigate the reinforcement of gender norms and promote equitable participation, the visualization of family members' efforts could be designed to emphasize collective efforts, or a combination of collective and individual contributions.

\subsection{Enhancing Parental Involvement in Daily Family Learning}

Our findings reveal that \tool{} contributes to improving parental collaboration through task allocation and shared goals, shifting from individual efforts to more collaborative family practices (Section~\ref{finding1}).
According to Epstein’s Six Types of Involvement framework \cite{epstein2018school}, increased parental participation positively impacts student performance and strengthens the partnership between schools and parents.

Consistent with the principle of learning at home in Epstein’s framework, our findings highlight the importance of involving more caregivers in everyday family learning activities, such as tutoring homework, engaging in interactive activities (e.g., sports), and cultivating life skills. Prior studies also suggest that parents want opportunities to be involved and to monitor their children’s learning~\cite{garg2020conversational, zhang2022storybuddy}. These shared responsibilities not only support children’s development but also increase caregivers’ awareness of the hidden labor involved in family work, fostering mutual understanding among caregivers. However, rather than emphasizing only the school-family partnership, our findings contribute by expanding this perspective to highlight the importance of communication and coordination within the family learning context. This contribution underscores that collaboration among caregivers is a foundation for supporting children’s learning effectively.

As for other dimensions, and aligning with Epstein’s definition of parenting, this refers to all the activities parents engage in to raise happy, healthy children who become capable students. Rather than merely focusing on academic content, the activities supported by \tool{} promote the development of children’s life skills, cultural literacy, and physical well-being. This is achieved by breaking down abstract parenting goals into concrete, age-appropriate subtasks that caregivers can collaboratively integrate into daily family learning. Moreover, by scaffolding the tutoring process through actionable and concrete task suggestions, \tool{} promotes informal ``volunteering'' among less active caregivers, such as non-primary caregivers, thereby encouraging collaborative engagement in the family learning context. These insights provide important implications for the design of AI-mediated tools in family learning. Specifically, by lowering participation thresholds and offering more concrete task descriptions, such design strategies have the potential to distribute responsibilities more evenly and enhance the sense of shared responsibility among caregivers.

\subsection{Limitations}

This study has several limitations. 
First, children were not included as participants, so our findings are based primarily on caregivers’ perspectives. As a result, we may not fully capture children’s experiences, preferences, and needs when interacting with AI learning tools. Future research should engage children directly, for example by involving them as co-designers in family learning contexts.
Second, our work focused only on caregivers, without considering other stakeholders such as teachers or schools. Yet, family learning is often embedded in a broader educational ecosystem that connects home and school. Future studies could explore how AI-powered tools support collaboration across these settings, bridging formal and informal learning opportunities. 
Third, we did not examine the use of \tool{} in multi-child families. Coordination and responsibility-sharing can be more complex when multiple children are involved, and different design considerations may be needed to support such contexts. Investigating these dynamics will be important for extending the applicability of AI-mediated family learning tools.

\section{Conclusion} 

This study examines how AI-empowered learning tools can support collaboration among caregivers in daily family learning. From a formative study, we identified four key insights: expertise- and availability-based task division, limited caregiver collaboration, imbalanced division of labor, and parental mediation of children’s AI tool use. Guided by these insights, we designed \tool, an AI system that scaffolds collaboration by distributing tasks, surfacing hidden contributions, and providing tutoring support. Findings from a field study with 11 families show that AI tools can facilitate both caregiver-child and caregiver-caregiver collaboration through structured, age-appropriate tasks and shared responsibilities, while also revealing challenges in capturing incidental learning opportunities and accounting for caregiving effort.
This work contributes to a broader understanding of how AI can be designed not only to support children’s learning but also to foster more coordinated and sustainable caregiving practices in home environments.


\begin{thebibliography}{67}
\providecommand{\natexlab}[1]{#1}
\providecommand{\url}[1]{\texttt{#1}}
\expandafter\ifx\csname urlstyle\endcsname\relax
  \providecommand{\doi}[1]{doi: #1}\else
  \providecommand{\doi}{doi: \begingroup \urlstyle{rm}\Url}\fi

\bibitem[Braun and Clarke(2006)]{braun2006using}
Virginia Braun and Victoria Clarke.
\newblock Using thematic analysis in psychology.
\newblock \emph{Qualitative research in psychology}, 3\penalty0 (2):\penalty0 77--101, 2006.

\bibitem[Brown et~al.(2007)Brown, Taylor, Izadi, Sellen, Kaye, and Eardley]{brown2007locating}
Barry Brown, Alex~S Taylor, Shahram Izadi, Abigail Sellen, Joseph~Jofish’ Kaye, and Rachel Eardley.
\newblock Locating family values: A field trial of the whereabouts clock: (nominated for the best paper award).
\newblock In \emph{UbiComp 2007: Ubiquitous Computing: 9th International Conference, UbiComp 2007, Innsbruck, Austria, September 16-19, 2007. Proceedings 9}, pages 354--371. Springer, 2007.

\bibitem[Chen et~al.(2025)Chen, Tang, Lu, Yao, Fan, Ma, Xu, Wang, Sun, and He]{chen2025characterizing}
Jiaju Chen, Minglong Tang, Yuxuan Lu, Bingsheng Yao, Elissa Fan, Xiaojuan Ma, Ying Xu, Dakuo Wang, Yuling Sun, and Liang He.
\newblock Characterizing llm-empowered personalized story reading and interaction for children: Insights from multi-stakeholder perspectives.
\newblock In \emph{Proceedings of the 2025 CHI Conference on Human Factors in Computing Systems}, pages 1--24, 2025.

\bibitem[Davidoff et~al.(2010)Davidoff, Zimmerman, and Dey]{davidoff2010routine}
Scott Davidoff, John Zimmerman, and Anind~K. Dey.
\newblock How routine learners can support family coordination.
\newblock In \emph{Proceedings of the SIGCHI Conference on Human Factors in Computing Systems}, pages 2461--2470. ACM, 2010.

\bibitem[DiMicco et~al.(2004)DiMicco, Pandolfo, and Bender]{dimicco2004influencing}
Joan~Morris DiMicco, Anna Pandolfo, and Walter Bender.
\newblock Influencing group participation with a shared display.
\newblock In \emph{Proceedings of the 2004 ACM conference on Computer supported cooperative work}, pages 614--623, 2004.

\bibitem[Druga et~al.(2022)Druga, Christoph, and Ko]{druga2022family}
Stefania Druga, Fee~Lia Christoph, and Amy~J Ko.
\newblock Family as a third space for ai literacies: How do children and parents learn about ai together?
\newblock In \emph{Proceedings of the 2022 CHI conference on human factors in computing systems}, pages 1--17, 2022.

\bibitem[{Duolingo}(2025)]{duolingo2025}
{Duolingo}.
\newblock Duolingo ai tutor.
\newblock \url{https://www.duolingo.com}, 2025.
\newblock Accessed: 2025-09-06.

\bibitem[Dworkin et~al.(2016)Dworkin, McCann, and McGuire]{dworkin2016coparenting}
Jodi Dworkin, Ellie McCann, and Jenifer~K McGuire.
\newblock Coparenting in the digital era: Exploring divorced parents’ use of technology.
\newblock In \emph{Divorce, separation, and remarriage: The transformation of family}. Emerald Group Publishing Limited, 2016.

\bibitem[Eira et~al.(2025)Eira, Rasouli, and Charisi]{eira2025parents}
Maria Eira, Amirkaveh Rasouli, and Vicky Charisi.
\newblock Parents’ perceptions about the use of generative ai systems by adolescents.
\newblock In \emph{Proceedings of the 24th Interaction Design and Children}, pages 927--931. 2025.

\bibitem[Elwahsh et~al.(2025)Elwahsh, Stern, Singh, and Ayobi]{elwahsh2025linguistic}
Sarah Elwahsh, Nora Stern, Aneesha Singh, and Amid Ayobi.
\newblock Linguistic diversity and mental well-being: Co-designing custom ai chatbots with multilingual mothers.
\newblock In \emph{Proceedings of the 7th ACM Conference on Conversational User Interfaces}, pages 1--17, 2025.

\bibitem[Epstein(2018)]{epstein2018school}
Joyce~L. Epstein.
\newblock \emph{School, Family, and Community Partnerships: Preparing Educators and Improving Schools}.
\newblock Routledge, 2018.

\bibitem[Falk and Dierking(2000)]{falk2000learning}
John~H. Falk and Lynn~D. Dierking.
\newblock \emph{Learning from Museums: Visitor Experiences and the Making of Meaning}.
\newblock AltaMira Press, 2000.

\bibitem[Gao et~al.(2025)Gao, Liu, Tang, Liu, Yu, Huang, Wang, Salim, Xu, Wei, et~al.]{gao2025homework}
Nan Gao, Yibin Liu, Xin Tang, Yanyan Liu, Chun Yu, Yun Huang, Yuntao Wang, Flora~D Salim, Xuhai Xu, Jun Wei, et~al.
\newblock The homework wars: Exploring emotions, behaviours, and conflicts in parent-child homework interactions.
\newblock \emph{Proceedings of the ACM on Interactive, Mobile, Wearable and Ubiquitous Technologies}, 9\penalty0 (3):\penalty0 1--37, 2025.

\bibitem[Garg and Sengupta(2020)]{garg2020conversational}
Radhika Garg and Subhasree Sengupta.
\newblock Conversational technologies for in-home learning: using co-design to understand children's and parents' perspectives.
\newblock In \emph{Proceedings of the 2020 CHI conference on human factors in computing systems}, pages 1--13, 2020.

\bibitem[Grant et~al.(2016)Grant, Rodger, and Hoffmann]{grant2016intervention}
N~Grant, S~Rodger, and T~Hoffmann.
\newblock Intervention decision-making processes and information preferences of parents of children with autism spectrum disorders.
\newblock \emph{Child: care, health and development}, 42\penalty0 (1):\penalty0 125--134, 2016.

\bibitem[Han and Han(2025)]{han2025empowering}
Ariel Han and Shenshen Han.
\newblock Empowering children’s ai literacy through co-creating stories with llm.
\newblock In \emph{Proceedings of the 24th Interaction Design and Children}, pages 994--999. 2025.

\bibitem[Han et~al.(2024)Han, Zhou, Cai, Han, Ko, Corrigan, and Peppler]{han2024teachers}
Ariel Han, Xiaofei Zhou, Zhenyao Cai, Shenshen Han, Richard Ko, Seth Corrigan, and Kylie~A Peppler.
\newblock Teachers, parents, and students' perspectives on integrating generative ai into elementary literacy education.
\newblock In \emph{Proceedings of the 2024 CHI Conference on Human Factors in Computing Systems}, pages 1--17, 2024.

\bibitem[Ho(2025)]{ho2025empowering}
Hui-Ru Ho.
\newblock Empowering parents and teachers to support children's learning through ai-based and robotic learning companions.
\newblock In \emph{Proceedings of the Extended Abstracts of the CHI Conference on Human Factors in Computing Systems}, pages 1--5, 2025.

\bibitem[Ho et~al.(2025)Ho, Kargeti, Liu, and Mutlu]{ho2025set}
Hui-Ru Ho, Nitigya Kargeti, Ziqi Liu, and Bilge Mutlu.
\newblock Set-paired: Designing for parental involvement in learning with an ai-assisted educational robot.
\newblock In \emph{Proceedings of the 2025 CHI Conference on Human Factors in Computing Systems}, pages 1--20, 2025.

\bibitem[Hochschild(1983)]{hochschild1983managedheart}
Arlie~Russell Hochschild.
\newblock \emph{The Managed Heart: Commercialization of Human Feeling}.
\newblock University of California Press, 1983.

\bibitem[Huang et~al.(2025)Huang, Lo, Liu, Luo, and Jin]{huang2025toward}
Gaojian Huang, Wei-Hsiang Lo, Guannan Liu, Yue Luo, and Yantong Jin.
\newblock Toward ai standardization: A triadic human-ai collaboration framework for multi-level autonomous mobility.
\newblock In \emph{2025 IEEE Conference on Artificial Intelligence (CAI)}, pages 1568--1577. IEEE, 2025.

\bibitem[Jo et~al.(2020)Jo, Toombs, Gray, and Hong]{jo2020parentingstress}
Eunkyung Jo, Austin~L. Toombs, Colin~M. Gray, and Hwajung Hong.
\newblock Understanding parenting stress through co-designed self-trackers.
\newblock In \emph{Proceedings of the 2020 CHI Conference on Human Factors in Computing Systems}, pages 1--13. ACM, 2020.

\bibitem[Kelly et~al.(2016)Kelly, Watts, and Payne]{kelly2016can}
Ryan Kelly, Leon Watts, and Stephen~J Payne.
\newblock Can visualization of contributions support fairness in collaboration? findings from meters in an online game.
\newblock In \emph{Proceedings of the 19th ACM Conference on Computer-Supported Cooperative Work \& Social Computing}, pages 664--678, 2016.

\bibitem[{Khan Academy}(2025)]{khanmigo2025}
{Khan Academy}.
\newblock Khanmigo ai tutor.
\newblock \url{https://www.khanmigo.ai}, 2025.
\newblock Accessed: 2025-09-06.

\bibitem[Kim et~al.(2025)Kim, Sato, White, Ho, Lee, Hwang, and Mutlu]{kim2025bridging}
Callie~Y. Kim, Arissa~J. Sato, Nathan~T. White, Hui-Ru Ho, Christine~P. Lee, Yuna Hwang, and Bilge Mutlu.
\newblock Bridging generations using ai-supported co-creative activities.
\newblock In \emph{Proceedings of the 2025 CHI Conference on Human Factors in Computing Systems}, pages 1--15. ACM, 2025.

\bibitem[Kirchner et~al.(2020)Kirchner, Sakaguchi-Tang, Michelson, Munson, and Kientz]{kirchner2020just}
Susanne Kirchner, Dawn~K Sakaguchi-Tang, Rebecca Michelson, Sean~A Munson, and Julie~A Kientz.
\newblock "this just felt to me like the right thing to do" decision-making experiences of parents of young children.
\newblock In \emph{Proceedings of the 2020 ACM designing interactive systems conference}, pages 489--503, 2020.

\bibitem[Kurian(2025)]{kurian2025designing}
Nomisha Kurian.
\newblock Designing child-safe conversational ai: Three dilemmas for responsible design.
\newblock In \emph{Proceedings of the 7th ACM Conference on Conversational User Interfaces}, pages 1--5, 2025.

\bibitem[Lin et~al.(2024)Lin, Li, Huang, Ecsedy, Feinberg, Teti, and Carroll]{lin2024coparenting}
Ya-Fang Lin, Na~Li, Wan-Hsuan Huang, Karen Ecsedy, Mark~E. Feinberg, Douglas Teti, and John~M. Carroll.
\newblock “ultimately we’re together”: Understanding new parents’ experiences of co-parenting.
\newblock \emph{Proceedings of the ACM on Human-Computer Interaction}, 8\penalty0 (CSCW2):\penalty0 1--25, 2024.

\bibitem[Lin et~al.(2025)Lin, Li, Huang, Prabavathi, Cai, and Carroll]{lin2025parental}
Ya-Fang Lin, Xiaotian Li, Wan-Hsuan Huang, Charan~Pushpanathan Prabavathi, Jie Cai, and John~M Carroll.
\newblock Parental collaboration and closeness: Envisioning with new couple parents.
\newblock In \emph{Proceedings of the 2025 ACM Designing Interactive Systems Conference}, pages 2637--2651, 2025.

\bibitem[Liu et~al.(2025)Liu, Zha, Zhang, Wang, Zhang, Xin, Nie, Zhang, and Xu]{liu2025bricksmart}
Yujia Liu, Siyu Zha, Yuewen Zhang, Yanjin Wang, Yangming Zhang, Qi~Xin, Lun~Yiu Nie, Chao Zhang, and Yingqing Xu.
\newblock Bricksmart: Leveraging generative ai to support children's spatial language learning in family block play.
\newblock In \emph{Proceedings of the 2025 CHI Conference on Human Factors in Computing Systems}, pages 1--19, 2025.

\bibitem[Long et~al.(2022)Long, Teachey, and Magerko]{long2022family}
Duri Long, Anthony Teachey, and Brian Magerko.
\newblock Family learning talk in ai literacy learning activities.
\newblock In \emph{Proceedings of the 2022 CHI Conference on Human Factors in Computing Systems}, pages 1--20, 2022.

\bibitem[Moll et~al.(1992)Moll, Amanti, Neff, and Gonzalez]{moll1992funds}
Luis~C. Moll, Cathy Amanti, Deborah Neff, and Norma Gonzalez.
\newblock Funds of knowledge for teaching: Using a qualitative approach to connect homes and classrooms.
\newblock In \emph{Theory into Practice}, volume~31, pages 132--141. Taylor \& Francis, 1992.

\bibitem[Neustaedter and Bernheim~Brush(2006)]{neustaedter2006linc}
Carman Neustaedter and AJ~Bernheim~Brush.
\newblock "linc-ing" the family: the participatory design of an inkable family calendar.
\newblock In \emph{Proceedings of the SIGCHI conference on Human Factors in computing systems}, pages 141--150, 2006.

\bibitem[Neustaedter et~al.(2007)Neustaedter, Brush, and Greenberg]{neustaedter2007digital}
Carman Neustaedter, AJ~Bernheim Brush, and Saul Greenberg.
\newblock A digital family calendar in the home: lessons from field trials of linc.
\newblock In \emph{Proceedings of Graphics interface 2007}, pages 199--20, 2007.

\bibitem[Neustaedter et~al.(2009)Neustaedter, Brush, and Greenberg]{neustaedter2009calendar}
Carman Neustaedter, AJ~Bernheim Brush, and Saul Greenberg.
\newblock The calendar is crucial: Coordination and awareness through the family calendar.
\newblock \emph{ACM Transactions on Computer-Human Interaction (TOCHI)}, 16\penalty0 (1):\penalty0 1--48, 2009.

\bibitem[Nguyen et~al.(2024)Nguyen, Borchers, Xia, and Aleven]{nguyen2024designing}
Ha~Tien Nguyen, Conrad Borchers, Meng Xia, and Vincent Aleven.
\newblock Designing tools for caregiver involvement in intelligent tutoring systems for middle school mathematics.
\newblock In \emph{Proceedings of the 18th International Conference of the Learning Sciences-ICLS 2024, pp. 865-872}, 2024.

\bibitem[Nikkhah et~al.(2022)Nikkhah, John, Yalamarti, Mueller, and Miller]{nikkhah2022family}
Sarah Nikkhah, Swaroop John, Krishna~Supradeep Yalamarti, Emily~L Mueller, and Andrew~D Miller.
\newblock Family care coordination in the children's hospital: Phases and cycles in the pediatric cancer caregiving journey.
\newblock \emph{Proceedings of the ACM on Human-Computer Interaction}, 6\penalty0 (CSCW2):\penalty0 1--30, 2022.

\bibitem[Odom et~al.(2010)Odom, Zimmerman, and Forlizzi]{odom2010designing}
William Odom, John Zimmerman, and Jodi Forlizzi.
\newblock Designing for dynamic family structures: divorced families and interactive systems.
\newblock In \emph{Proceedings of the 8th ACM conference on designing interactive systems}, pages 151--160, 2010.

\bibitem[Pan et~al.(2015)Pan, Forghani, Neustaedter, Strauss, and Guindon]{pan2015family}
Rui Pan, Azadeh Forghani, Carman Neustaedter, Nick Strauss, and Ashley Guindon.
\newblock The family board: An information sharing system for family members.
\newblock In \emph{Proceedings of the 18th acm conference companion on computer supported cooperative work \& social computing}, pages 207--210, 2015.

\bibitem[Perella-Holfeld et~al.(2024)Perella-Holfeld, Sallam, Petrie, Gomez, Irani, and Sakamoto]{perella2024parent}
Francisco Perella-Holfeld, Samar Sallam, Julia Petrie, Randy Gomez, Pourang Irani, and Yumiko Sakamoto.
\newblock Parent and educator concerns on the pedagogical use of ai-equipped social robots.
\newblock \emph{Proceedings of the ACM on Interactive, Mobile, Wearable and Ubiquitous Technologies}, 8\penalty0 (3):\penalty0 1--34, 2024.

\bibitem[Pina et~al.(2017)Pina, Sien, Ward, Yip, Munson, Fogarty, and Kientz]{pina2017personal}
Laura~R Pina, Sang-Wha Sien, Teresa Ward, Jason~C Yip, Sean~A Munson, James Fogarty, and Julie~A Kientz.
\newblock From personal informatics to family informatics: Understanding family practices around health monitoring.
\newblock In \emph{Proceedings of the 2017 acm conference on computer supported cooperative work and social computing}, pages 2300--2315, 2017.

\bibitem[Quan et~al.(2024)Quan, Du, and Ding]{quan2024young}
Shuang Quan, Yao Du, and Yi~Ding.
\newblock Young children and chatgpt: Parents' use of chatgpt in parenting.
\newblock In \emph{Extended Abstracts of the CHI Conference on Human Factors in Computing Systems}, pages 1--7, 2024.

\bibitem[Quan et~al.(2025)Quan, Du, and Lyu]{quan2025parents}
Shuang Quan, Yao Du, and Yao Lyu.
\newblock Parents, children, and chatgpt in home environments: The conversation content and the interaction mode.
\newblock In \emph{Proceedings of the Extended Abstracts of the CHI Conference on Human Factors in Computing Systems}, pages 1--6, 2025.

\bibitem[{Quark}(2024)]{quarkgaokao2024}
{Quark}.
\newblock Quark gaokao application system.
\newblock \url{https://gaokao.quark.cn}, 2024.
\newblock Accessed: 2025-09-06.

\bibitem[Reitberger et~al.(2009)Reitberger, M{\"o}ller, Kranz, Alt, Rukzio, and Schmidt]{reitberger2009tasks}
Wolfgang Reitberger, Andreas M{\"o}ller, Matthias Kranz, Florian Alt, Enrico Rukzio, and Albrecht Schmidt.
\newblock Task sharing and appreciation in families: Exploring the design of cooperative technologies.
\newblock In \emph{Proceedings of the 3rd International Conference on Tangible and Embedded Interaction}, pages 67--74. ACM, 2009.

\bibitem[Reitberger et~al.(2013)Reitberger, Kastenmiller, and Fitzpatrick]{reitberger2013invisible}
Wolfgang Reitberger, Martin Kastenmiller, and Geraldine Fitzpatrick.
\newblock Invisible work: an ambient system for awareness and reflection of household tasks.
\newblock In \emph{Persuasive Technology: 8th International Conference, PERSUASIVE 2013, Sydney, NSW, Australia, April 3-5, 2013. Proceedings 8}, pages 180--191. Springer, 2013.

\bibitem[Rogoff(2003)]{rogoff2003cultural}
Barbara Rogoff.
\newblock \emph{The Cultural Nature of Human Development}.
\newblock Oxford University Press, 2003.

\bibitem[Rosman et~al.(2025)Rosman, Suneesh, and Yadollahi]{rosman2025exploring}
Amirah Nabilah~Binti Rosman, Shyamli Suneesh, and Elmira Yadollahi.
\newblock Exploring parental ai literacy and perceptions of robot transparency in educational child-robot interaction.
\newblock In \emph{Proceedings of the 24th Interaction Design and Children}, pages 984--988. 2025.

\bibitem[Sannon et~al.(2020)Sannon, Vorvoreanu, Walker, and Fourney]{sannon2020doing}
Shruti Sannon, Mihaela Vorvoreanu, Kathleen Walker, and Adam Fourney.
\newblock "am i doing this all wrong?" challenges and opportunities in family information management.
\newblock \emph{Proceedings of the ACM on Human-Computer Interaction}, 4\penalty0 (CSCW2):\penalty0 1--28, 2020.

\bibitem[Schaefbauer et~al.(2015)Schaefbauer, Khan, Le, Sczechowski, and Siek]{schaefbauer2015snack}
Christopher~L Schaefbauer, Danish~U Khan, Amy Le, Garrett Sczechowski, and Katie~A Siek.
\newblock Snack buddy: supporting healthy snacking in low socioeconomic status families.
\newblock In \emph{Proceedings of the 18th acm conference on computer supported cooperative work \& social computing}, pages 1045--1057, 2015.

\bibitem[Seo et~al.(2025)Seo, Kim, Kim, Fan, Ackerman, Choi, and Park]{seo2025enhancing}
Woosuk Seo, Young-Ho Kim, Ji~Eun Kim, Megan~Tao Fan, Mark~S Ackerman, Sung~Won Choi, and Sun~Young Park.
\newblock Enhancing pediatric communication: The role of an ai-driven chatbot in facilitating child-parent-provider interaction.
\newblock In \emph{Proceedings of the 2025 CHI Conference on Human Factors in Computing Systems}, pages 1--16, 2025.

\bibitem[Shang et~al.(2025)Shang, Wang, and Qi]{shang2025learner}
Yi~Shang, Jingyun Wang, and Xiaofei Qi.
\newblock A learner-ai-parent collaboration framework for home learning environment.
\newblock In \emph{International Conference on Artificial Intelligence in Education}, pages 359--364, 2025.

\bibitem[Shin et~al.(2022)Shin, Peng, and Lee]{shin2022more}
Ji~Youn Shin, Wei Peng, and Hee~Rin Lee.
\newblock More than bedtime and the bedroom: Sleep management as a collaborative work for the family.
\newblock In \emph{CHI Conference on Human Factors in Computing Systems}, pages 1--16, 2022.

\bibitem[Sohn et~al.(2012)Sohn, Lee, Zhang, Dearman, and Truong]{sohn2012examination}
Timothy Sohn, Lorikeet Lee, Stephanie Zhang, David Dearman, and Khai Truong.
\newblock An examination of how households share and coordinate the completion of errands.
\newblock In \emph{Proceedings of the ACM 2012 conference on Computer Supported Cooperative Work}, pages 729--738, 2012.

\bibitem[Song et~al.(2018)Song, Kim, Kang, Park, and Kim]{song2018bebecode}
Seokwoo Song, Juho Kim, Bumsoo Kang, Wonjeong Park, and John Kim.
\newblock Bebecode: Collaborative child development tracking system.
\newblock In \emph{Proceedings of the 2018 chi conference on human factors in computing systems}, pages 1--12, 2018.

\bibitem[Sun et~al.(2024)Sun, Chen, Yao, Liu, Wang, Ma, Lu, Xu, and He]{sun2024exploring}
Yuling Sun, Jiaju Chen, Bingsheng Yao, Jiali Liu, Dakuo Wang, Xiaojuan Ma, Yuxuan Lu, Ying Xu, and Liang He.
\newblock Exploring parent's needs for children-centered ai to support preschoolers' interactive storytelling and reading activities.
\newblock \emph{Proceedings of the ACM on Human-Computer Interaction}, 8\penalty0 (CSCW2):\penalty0 1--25, 2024.

\bibitem[Tausch et~al.(2016)Tausch, Ta, and Hussmann]{tausch2016comparison}
Sarah Tausch, Stephanie Ta, and Heinrich Hussmann.
\newblock A comparison of cooperative and competitive visualizations for co-located collaboration.
\newblock In \emph{Proceedings of the 2016 CHI conference on human factors in computing systems}, pages 5034--5039, 2016.

\bibitem[Taylor et~al.(2006)Taylor, Swan, Eardley, Sellen, Hodges, and Wood]{taylor2006augmenting}
Alex~S Taylor, Laurel Swan, Rachel Eardley, Abigail Sellen, Steve Hodges, and Ken Wood.
\newblock Augmenting refrigerator magnets: why less is sometimes more.
\newblock In \emph{Proceedings of the 4th Nordic conference on Human-computer interaction: changing roles}, pages 115--124, 2006.

\bibitem[Venugopalan et~al.(2025)Venugopalan, Yan, Borchers, Lin, and Aleven]{venugopalan2025combining}
Devika Venugopalan, Ziwen Yan, Conrad Borchers, Jionghao Lin, and Vincent Aleven.
\newblock Combining large language models with tutoring system intelligence: A case study in caregiver homework support.
\newblock In \emph{Proceedings of the 15th International Learning Analytics and Knowledge Conference}, pages 373--383, 2025.

\bibitem[Wang et~al.(2023)Wang, Peng, Que, Liu, Zhou, Wu, Guo, Gan, Ni, Yang, et~al.]{wang2023rolellm}
Zekun~Moore Wang, Zhongyuan Peng, Haoran Que, Jiaheng Liu, Wangchunshu Zhou, Yuhan Wu, Hongcheng Guo, Ruitong Gan, Zehao Ni, Jian Yang, et~al.
\newblock Rolellm: Benchmarking, eliciting, and enhancing role-playing abilities of large language models.
\newblock \emph{arXiv preprint arXiv:2310.00746}, 2023.

\bibitem[xiaoyuan souti.(2025)]{xiaoyuandouti2025}
xiaoyuan souti.
\newblock Xiaoyuan souti app.
\newblock \url{https://www.kanyun.com/}, 2025.
\newblock Accessed: 2025-09-06.

\bibitem[Yang(2025)]{yang2025home}
Yan Yang.
\newblock Are home educational robots really effective? insights from systematic review.
\newblock \emph{AI in Human Sciences}, 1\penalty0 (1):\penalty0 1--28, 2025.

\bibitem[Yarosh et~al.(2016)Yarosh, Schoenebeck, Kothaneth, and Bales]{yarosh2016best}
Svetlana Yarosh, Sarita Schoenebeck, Shreya Kothaneth, and Elizabeth Bales.
\newblock "best of both worlds" opportunities for technology in cross-cultural parenting.
\newblock In \emph{Proceedings of the 2016 chi conference on human factors in computing systems}, pages 635--647, 2016.

\bibitem[Yu et~al.(2021)Yu, DeVore, and Roque]{yu2021parental}
Junnan Yu, Andrea DeVore, and Ricarose Roque.
\newblock Parental mediation for young children’s use of educational media: A case study with computational toys and kits.
\newblock In \emph{Proceedings of the 2021 CHI conference on human factors in computing systems}, pages 1--12, 2021.

\bibitem[Yu(2025)]{yu2025safeguarding}
Yaman Yu.
\newblock Safeguarding children in generative ai: Risk frameworks and parental control tools.
\newblock In \emph{Companion Proceedings of the 2025 ACM International Conference on Supporting Group Work}, pages 121--123, 2025.

\bibitem[Zhang et~al.(2022)Zhang, Xu, Wang, Yao, Ritchie, Wu, Yu, Wang, and Li]{zhang2022storybuddy}
Zheng Zhang, Ying Xu, Yanhao Wang, Bingsheng Yao, Daniel Ritchie, Tongshuang Wu, Mo~Yu, Dakuo Wang, and Toby Jia-Jun Li.
\newblock Storybuddy: A human-ai collaborative chatbot for parent-child interactive storytelling with flexible parental involvement.
\newblock In \emph{Proceedings of the 2022 CHI Conference on Human Factors in Computing Systems}, pages 1--21, 2022.

\bibitem[Zuoyebang.(2025)]{zuoyebang2025}
Zuoyebang.
\newblock Zuoyebang website.
\newblock \url{https://www.zuoyebang.com/}, 2025.
\newblock Accessed: 2025-09-06.

\end{thebibliography}

\appendix
\end{document}